\documentclass[twocolumn,showpacs,amsmath,amstex,amssymb,mathfonts,prx]{revtex4-1}
\usepackage{amsthm,amsfonts,graphicx,verbatim,times}
\usepackage{amsmath}
\usepackage{amssymb}
\usepackage{amsthm}
\usepackage{amsfonts}
\usepackage{listings}
\usepackage{epstopdf}
\usepackage{color}
\usepackage{enumerate}
\usepackage{latexsym}
\usepackage[normalem]{ulem}
\usepackage{cancel}
\usepackage{hyperref}

\usepackage{psfrag}
\usepackage{bm}
\usepackage{graphicx}
\usepackage{subfigure}
\newcommand{\be}{\begin{equation}}
\newcommand{\ee}{\end{equation}}
\newcommand{\bea}{\begin{eqnarray}}
\newcommand{\eea}{\end{eqnarray}}

\renewcommand{\phi}{\varphi}
\renewcommand{\epsilon}{\varepsilon}

\begin{document}

\title{Solvable models for unitary and non-unitary topological phases}

\author{Z. Papi\'c$^{1,2}$}
\affiliation{$^1$ Perimeter Institute for Theoretical Physics, Waterloo, ON N2L 2Y5, Canada} 
\affiliation{$^2$ Institute for Quantum Computing, Waterloo, ON N2L 3G1, Canada}


\begin{abstract}
We introduce a broad class of simple models for quantum Hall states based on the expansion of their parent Hamiltonians near the one-dimensional limit of ``thin cylinders", i.e., when one dimension $L_y$ of the Hall surface becomes comparable to the magnetic length $\ell_B$. Formally, the models can be viewed as topological generalizations of the 1D Hubbard model with center-of-mass-preserving hopping of multiparticle clusters. In some cases, we show that the models can be exactly solved using elementary techniques, and yield simple wave functions for the ground states as well as the entire neutral excitation spectrum. We study a large class of Abelian and non-Abelian states in this limit, including the Read-Rezayi $\mathbb{Z}_k$ series, as well as states deriving from non-unitary or irrational conformal field theories -- the  ``Gaffnian", ``Haffnian", Haldane-Rezayi, and the ``permanent" state. We find that the thin-cylinder limit of unitary (rational) states is ``classical": their effective Hamiltonians reduce to only Hartree-type terms, the ground states are trivial insulators, and excitation gaps result from simple electrostatic repulsion. In contrast, for states deriving from non-unitary or irrational conformal field theories, the thin-cylinder limit is found to be intrinsically \emph{quantum} -- it contains hopping terms that play an important role in the structure of the ground states and in the energetics of the low-lying neutral excitations.  
\end{abstract}

\date{\today}

\maketitle 

\section{Introduction} \label{sec:introduction}

Partially filled Landau levels have long served as an important case study for the interplay of strong interactions with band topology, in particular through the realization of incompressible Fractional Quantum Hall (FQH) fluids~\cite{Tsui-PhysRevLett.48.1559}. Quite unexpectedly, much of theoretical progress has been made by formulating first-quantized many-body trial wave functions to describe these phases, and subsequently verifying their relevance for realistic systems in unbiased finite-size numerical calculations. Many of such wave functions can be written down with the help of conformal field theory (CFT)~\cite{Fubini-1991MPLA....6..347F,Moore1991362}. These include the celebrated Laughlin~\cite{Laughlin-PhysRevLett.50.1395} state, the Moore-Read ``Pfaffian"~\cite{Moore1991362} state and a more general Read-Rezayi (RR) sequence~\cite{Read-PhysRevB.59.8084}. 
The aforementioned states are good candidates to describe the low-energy physics of the incompressible (gapped) quantum fluids, experimentally manifested through the quantized plateau in the Hall conductance. The quasiparticle excitations of these states are very different -- Abelian or non-Abelian anyons -- depending on the state and its associated CFT. In order to understand the properties of these states, e.g., whether they have a gap for creating charge-neutral excitations, usually requires elaborate arguments based on mappings to screening plasmas~\cite{Laughlin-PhysRevLett.50.1395,Laughlin-PhysRevLett.50.1395,Bonderson-PhysRevB.83.075303}, which remain restricted to a few individual cases.

Therefore, it is desirable to find simpler models for quantum Hall states that are more tractable, yet reproduce their fundamental physical properties.
One of the goals of present work is to show how to generically construct such models for a large family of quantum Hall states by considering their parent ``pseudopotential" \cite{Haldane-PhysRevLett.51.605} Hamiltonians that yield the trial quantum Hall wave functions as unique zero modes, $H\Psi=0$. Parent Hamiltonians are available for many states (the notable exception being the composite fermion states~\cite{Jain:1989p294}, not considered here), and generally involve interactions between many-particle clusters. In all the cases, even if the exact zero-mode ground state of the Hamiltonian is known, the excited states of the pseudopotential Hamiltonian spectrum may not be solvable, and analytic insight could appear impossible. 

However, in constrast to the usual condensed matter systems defined on a lattice, the FQH systems are defined on a continuum 2D surface whose shape can be continuously varied. Hence, the FQH Hamiltonians have an important feature of the parametrical dependence of their matrix elements on the shape of the surface. As we show below, this allows to perform a reduction of $H$, nominally defined in two spatial dimensions, to the nearly ``one-dimensional limit", which is more tractable (and often exactly solvable), yielding a simple way to characterize not only the ground states, but also the entire neutral excitation spectrum. This limit includes the extreme 1D limit known as the ``thin-torus limit"~\cite{Bergholtz-PhysRevB.77.155308,Bergholtz-PhysRevLett.99.256803,Seidel-PhysRevLett.95.266405} in the literature, but also corrections to it when the aspect ratio of the Hall surface is slightly adjusted towards the isotropic (2D) limit.   

Formally, the dimensional reduction is accomplished in the Landau gauge~\cite{chak}, where the single-particle orbitals are chosen to be fully periodic along one cycle $L_y$, and Gaussian-localized in the other direction ($x$). Effectively, this corresponds to compactifying the Hall surface into a cylinder of perimeter $L_y$. Fully periodic boundary condition (torus) is also possible~\cite{chak}, and achieved by explicitly making the Gaussian part of the wave functions periodic in $L_x$, the cycle in $x$-direction. Setting the magnetic length to unity $\ell_B=1$, the separation of the Gaussians is controlled by the parameter $\kappa=2\pi/L_y$, while the width of each LL orbital is 1 (see Fig.~\ref{fig_cylinder}). Therefore, as one dimension ($L_y$) of the surface tends to zero, the overlap between the Gaussians vanishes, and the system starts to behave ``classically". This means that the deformation of the surface translates into a modification of the effective interaction between particles, effectively suppressing the off-diagonal Hamiltonian matrix elements, and leaving only the Hartree-type potential -- pure electrostatic repulsion. In this regime, referred to as the ``thin-torus limit" in the literature, the ground states are Tao-Thouless~\cite{Tao-PhysRevB.28.1142} (TT) crystals of particles (i.e., Slater determinants of electrons or permanents for bosons) pinned at Landau-gauge sites. Many early works have attempted to understand the nature of the FQH effect by utilizing such concepts~\cite{Anderson-PhysRevB.28.2264,Su-PhysRevB.30.1069,Su-PhysRevB.32.2617,Chui-PhysRevB.32.8438,Chui-PhysRevLett.56.2395}.

Although it is known that the thin-torus limit does not capture the physics of the FQH effect in all the details (e.g., the density of the ground state is not fully uniform but has charge-density-wave-like oscillations, the value of the excitation gap is not the same as in the isotropic limit, etc.), remarkably enough given its simplicity, it does predict correctly some properties of the system, such as the quantum numbers of the (degenerate) ground state(s) and charges of  quasiparticles. For gapped FQH states, these properties are generally thought to be adiabatically maintained~\cite{Seidel-PhysRevLett.95.266405} towards the isotropic 2D limit, as previous work has shown for the Abelian hierarchy states~\cite{Bergholtz-PhysRevLett.99.256803} which include the Laughlin and Jain composite fermion states. The hierarchy series terminates at filling factor $\nu=1/2$, which undergoes a metal-insulator transition~\cite{Bergholtz-PhysRevLett.99.256803,Bergholtz-PhysRevB.77.155308} from a gapped TT state to a gapless state of neutral fermions as $\kappa$ is varied. The non-Abelian Moore-Read Pfaffian state was also analyzed in this limit~\cite{Bergholtz-PhysRevB.74.081308,Seidel-PhysRevLett.97.056804,Ardonne-1742-5468-2008-04-P04016,Seidel-PhysRevLett.101.196802}. 

More recently, the thin-torus analysis was extended~\cite{Halperin83,Haldane-PhysRevLett.60.956} to ``multicomponent" states such as the Halperin states~\cite{Halperin83, Seidel-PhysRevLett.101.036804} and Haldane-Rezayi state~\cite{Seidel-PhysRevB.84.085122}. In the latter case, Ref.~\onlinecite{Seidel-PhysRevB.84.085122} argued the existence of gapless excitations and gave their phenomenological description. On a different front, it was emphasized that the Laughlin parent Hamiltonian is exactly solvable near the limit of thin torus or cylinder~\cite{Jansen-2012JMP....53l3306J,Soule-PhysRevB.85.155116,Nakamura-PhysRevLett.109.016401,Wang-PhysRevB.87.245119}, and this property was used to construct a mapping to the effective spin chain models~\cite{Nakamura-PhysRevLett.109.016401,Bergholtz2011755,Wang-PhysRevB.86.155104}. Connections between the two-body thin-torus Hamiltonians and Richardson-Gaudin models in theory of superconductivity were elucidated in Ref.~\onlinecite{Ortiz-PhysRevB.88.165303}. Very recently, the formalism of clustered Hamiltonians and their thin torus limits has been applied to the lattice analogs of FQH states in the absence of magnetic field~\cite{Bernevig-2012arXiv1204.5682B, Budich-PhysRevB.88.035139}. In particular, Ref.~\onlinecite{thomale} gave a closely related construction of short-range clustering Hamiltonians for fractional Chern insulators.

In this work we extend the thin-torus methodology for a large class of quantum Hall model states defined by short-range many-body parent Hamiltonians, including the entire Read-Rezayi series of states~\cite{Read-PhysRevB.59.8084}, the spin-polarized non-unitary (Gaffnian~\cite{Simon-PhysRevB.75.075318}) and irrational (Haffnian~\cite{Read-PhysRevB.61.10267,green-10thesis}) states, as well as spinful non-unitary states such as Haldane-Rezayi~\cite{Seidel-PhysRevB.84.085122} and the permanent state~\cite{Read-PhysRevB.54.16864}. We provide simple models for all these states (in some cases exactly solvable) that represent generalizations of the Hubbard model where hopping involves multiparticle clusters. 

Another motivation behind the present study is to contrast the behavior in the thin-cylinder limit of unitary (rational) states with that of the non-unitary or irrational states. It is known that non-unitary CFTs~\cite{yellow} naturally lead to states which possess diverging correlators at the edge of a quantum Hall droplet (assuming the edge CFT to be identical to the one used to construct the bulk trial state). For this reason, it has been conjectured~\cite{Read-PhysRevB.79.245304} that non-unitary model states can only describe gapless phases, and not bulk-incompressible fluids.  However, in practice the non-unitary states are often deceivingly similar to the unitary ones. For example, an elegant spin-singlet state for half filling of a Landau level -- the Haldane-Rezayi state~\cite{Haldane-PhysRevLett.60.956} -- was initially proposed as a wave function for the quantized plateau at filling factor $\nu=5/2$. Similarly, the  ``Gaffnian" state~\cite{Simon-PhysRevB.75.075318}, deriving from the non-unitary $M(5,3)$ minimal model~\cite{yellow}, appears very closely related to the composite fermion $\nu=2/5$ state~\cite{jainbook}, with almost no discernible difference in finite-system representations. General considerations imply that many of such non-unitary states could represent critical points between stable phases~\cite{Read-PhysRevB.61.10267,green-10thesis}, nevertheless it would be desirable to have a precise, microscopic diagnostic that could distinguish between unitary and other types of states. Direct numerical calculations based on exact diagonalization, for example, have been of little use in resolving this matter because small finite droplets of non-unitary states tend to appear ``gapped", and extrapolations to infinite systems have been inconclusive (some numerical studies, however, have given hints that the Gaffnian state fails to screen in the quasihole sector~\cite{Herland-PhysRevB.85.024520,Herland-PhysRevB.87.075117}). Note that unitarity of a CFT is by no means a guarantee of gapfulness of a state: for example, Laughlin wave functions at low filling factors no longer describe gapped liquids but states with charge-density-wave order. Thus, one might wonder if any of the higher-order $k\geq 3$ Read-Rezayi states similarly become gapless, and thus fundamentally fail to represent incompressible fluids. 

Here we demonstrate on several examples that the thin-torus behavior of unitary rational states is different from the non-unitary or irrational ones. We show that Read-Rezayi $\mathbb{Z}_k$ states have ``classical" description near the thin-torus limit, i.e., their effective Hamiltonians reduce to only Hartree-type terms, ground states are trivial insulators, with excitation gaps resulting from simple electrostatic repulsion. Corrections that introduce quantum fluctuations are in some cases analytically computable for small but finite $L_y$. In contrast, the thin-torus limit of the non-unitary states is found to be intrinsically \emph{quantum}: it contains hopping terms that play a crucial role in the structure of topologically degenerate ground states, as well as the energetics of the low-lying (neutral) excited states. As we illustrate in a number of cases,  the solvable models introduced here might provide ways to distinguish between unitary and non-unitary/irrational states, as well as to construct approximate descriptions of FQH states in the formalism of ``matrix-product states"~\cite{Zaletel-PhysRevB.86.245305, Estienne-PhysRevB.87.161112}. 

In Sec.~\ref{sec:ham} (and Appendix~\ref{sec_app}) we introduce the problem, discuss the clustering Hamiltonians that define quantum Hall states and how to adapt them to periodic boundary conditions in an efficient way that illuminates their underlying structure. The structure of the Hamiltonians and their positive semidefinite property is discussed in detail in Appendix~\ref{sec:factorization}. Sec.~\ref{sec:ham} moreover provides a detailed outline of our approach and introduces the notation. In Sec.~\ref{sec:rr} we study in detail the solvable models for the Read-Rezayi states, in particular the Laughlin, Moore-Read and $\mathbb{Z}_3$ RR cases. Secs.~\ref{sec:nonun} and \ref{sec:spin} are dedicated to the non-unitary and irrational states, without and with spin. Our conclusions are presented in Sec.~\ref{sec:conc}. 

\section{Clustered Hamiltonians on the torus}\label{sec:ham}

In this section we give some technical preliminaries and introduction to the single-particle problem on the torus, many-body Hamiltonians studied in this paper, and fix our notation and conventions.

\subsection{Single particle problem} 

We consider an electron in a magnetic field and subject to periodic boundary conditions on a unit cell $L_x\times L_y$. This boundary condition is compatible with the Landau gauge where the single-particle states are fully periodic along $L_y$ and Gaussian-localized along $x$-axis. When we want to enforce periodicity along $x$, the Gaussian part of the wave function must be explicitly made periodic, which yields the Jacobi theta functions for single-particle states. For the gauge choice $\mathbf{A}=(0,Bx,0)$, the one-body states are given by
\begin{equation}
\phi_j(\mathbf{r}) = \frac{1}{\sqrt{L_y\sqrt{\pi}}} \sum_k e^{i(X_j+kL_x)y-(X_j+kL_x+x)^2/2},\label{eq:onebodywf}
\end{equation}
where $j=0,\ldots,N_\phi-1$ ($N_\phi$ is the number of flux quanta) and $X_j=2\pi j/L_y$. The fundamental magnetic translations in $x-$ and $y-$direction are defined by 
\begin{equation}
t_x\equiv \exp(i\frac{L_y}{N_\phi}R_x), \;  t_y\equiv \exp(i\frac{L_x}{N_\phi}R_y),\label{eq:txty}
\end{equation}
where $\mathbf{R}$ is the guiding-center coordinate~\cite{chak}. Their action on the one-body states is
\begin{equation}
t_x \phi_j = \exp(-i\frac{2\pi}{N_\phi}j) \phi_j, \; t_y \phi_j = \phi_{j+1 ({\rm mod} \; N_\phi)}.\label{eq:txtyonebody}
\end{equation}
Thus the one-body states are eigenstates of $t_x$, and the fundamental magnetic translations form a projective representation since $t_x t_y = t_y t_x e^{i2\pi/N_\phi}$. Their many-body extensions~\cite{Haldane-PhysRevLett.55.2095} can be used to classify the states of interacting particles, and deduce the minimal $q$-fold degeneracy inherent to every state at filling factor $\nu=N/N_\phi=p/q$ ($N$ is the number of particles).  

\begin{figure}[t]
\centerline{\includegraphics[width=\linewidth]{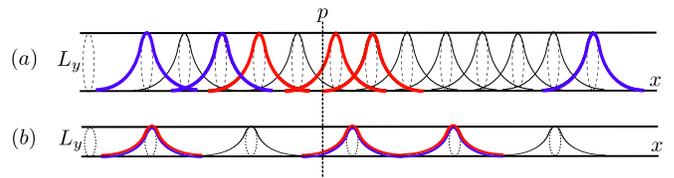}}
\caption[]{(Color online). Quantum Hall cylinder with a 1D chain of Landau-gauge orbitals, fully periodic along the perimeter $L_y$ and Gaussian-localized in $x$. The separation between adjacent orbitals is given by $\kappa=2\pi/L_y$, while the width of each orbital is $\ell_B=1$. In the isotropic (2D) limit (a), $L_y$ is large and orbitals strongly overlap, thus complicated long-range hopping processes become possible. For example, three particles might be destroyed in orbitals 4, 6, 7 (red), and created in orbitals 1, 3, 13 (blue). Only hoppings that preserve the center of mass $p$ are allowed. In the thin-cylinder limit (b), hoppings are suppressed, and the main terms in the Hamiltonian are density-density repulsions.}
\label{fig_cylinder}
\end{figure}

The limit of ``thin torus" formally corresponds to $L_y\to 0$ under the constraint that the total magnetic flux $N_\phi$ through the surface remains quantized $L_x L_y=2\pi N_\phi$. Thus, the parameter $\kappa=2\pi/L_y$ characterizing the overlap between one-body
orbitals becomes large (on a finite torus, this can be equivalently achieved by varying the aspect ratio $L_x/L_y$), and the individual matrix elements of the interaction tend to those in the cylinder geometry, Fig.~\ref{fig_cylinder}. In this work, we will consider both torus and finite cylinder geometry. 

\subsection{Many-body Hamiltonians on the torus}

A clustered Hamiltonian for bosons, of the Read-Rezayi $(k,r)$-type, penalizes (assigns positive energy) to a cluster of $k+1$ particles in $r$ consecutive orbitals, whereas there is no energy penalty for clusters of less than $k+1$ particles in $r$ orbitals. Under periodic boundary conditions, the system retains translational invariance, but rotational symmetry of the infinite plane, which underlies the formalism of projection Hamiltonians~\cite{Simon-PhysRevB.75.075318}, is no longer exact.  Instead, the short-distance clustering properties and projection Hamiltonians have to be formulated by combining delta functions and their derivatives, and making them appropriately periodic. For bosons, the primary Read-Rezayi series with $r=2$ is defined by the Hamiltonian
\begin{equation}
H_{(k,2)} = \sum_{i_1<\ldots <i_{k+1}}\delta^2(\mathbf{r}_{i_1}-\mathbf{r}_{i_2}) \ldots \delta^2(\mathbf{r}_{i_{k}}-\mathbf{r}_{i_{k+1}}).\label{RRHamiltonian}
\end{equation}
At particular filling factors $\nu=N/N_\phi=k/(k+2)$, these Hamiltonians possess densest zero-energy ground states on the torus which are $k+1$-degenerate. For example, the Laughlin state at $\nu=1/2$ is the (two-fold degenerate) ground-state of the contact interaction between two bosons, $\sum_{i<j}\delta(\mathbf{r}_i-\mathbf{r}_j)$. 

Pseudopotential Hamiltonians like Eq.~(\ref{RRHamiltonian}) are defined for an infinite system. On a finite torus, the interaction must be made periodic upon translating every particle coordinate by a multiple of $L_x$ or $L_y$. This is most easily achieved by considering the interaction Fourier transform, 
\begin{equation}
\sum_{\mathbf{q}_1,\ldots,\mathbf{q}_{k}} \tilde{V}(\mathbf{q}_1,\ldots,\mathbf{q}_{k}) e^{i\mathbf{q}_1(\mathbf{r_1}-\mathbf{r}_2)} \ldots e^{i\mathbf{q}_{k}(\mathbf{r}_{k}-\mathbf{r}_{k+1})},\label{eq:HamiltonianFourier}
\end{equation}
and assuming it to be defined on a Brillouin zone discrete mesh $\mathbf{q}_i = (2\pi s_i/L_x, 2\pi t_i/L_y)$, where $s_i,t_i$ are integers. On a cylinder, the $x$-coordinate is considered infinite, and thus sums over $q_x$ components of momenta can be converted into integrals.

Although correct in principle, the above approach of summing over discretized $\mathbf{q}$ momenta on the torus is not very illuminating and furthermore becomes extremely time-consuming in numerics due to the nested summations in Eq.~(\ref{eq:HamiltonianFourier}). Fortunately, it is possible to significantly reduce the amount of computation by performing an additional resummation over $q_{1x},\ldots,q_{k,x}$ using the Poisson summation formula, as shown in Appendix~\ref{sec_app}. This way, the matrix element describing the $k+1$-body process of scattering from states $j_1,\ldots,j_{k+1}$ into $j_{k+2},\ldots,j_{2k+2}$ on the cylinder factorizes into a product of two parts 
\begin{eqnarray}
V_{j_1,\ldots,j_{2k+2}}&=&\left\langle j_1,...,j_{k+1} |H_{(k,2)}  | j_{k+2},...,j_{2k+2} \right\rangle\nonumber\\
&=&\bar{f}(j_1,\ldots,j_{k+1}) f(j_{k+2},\ldots,j_{2k+2}),\label{eq:matrixelement}
\end{eqnarray}
where $\bar f$ represents part of the scattering amplitude that depends solely on the occupation numbers of states being created ($c_{j_1}^\dagger\ldots c_{j_{k+1}}^\dagger$), and $f$ depends on the annihilated states $c_{j_{k+2}}\ldots c_{j_{2k+2}}$. Such a factorization arises naturally in the symmetric gauge when FQH systems are studied on the disk or sphere geometry, and is crucial in understanding the clustering properties~\cite{Chandran-PhysRevB.84.205136} from the model Hamiltonians. 
On the torus, the Poisson formula generally admits to reorganize the $2k$ above sums (Eq.~\ref{eq:HamiltonianFourier}) in the following manner (see Appendix~\ref{sec_app}) 
\begin{eqnarray}\label{torusmatel}
\nonumber V_{j_1,\ldots,j_{2k+2}} &=& \sum_{g=0,\ldots,k} \{ \sum_{l_1,\ldots,l_{k+1}} \bar f(\tilde{j}_1,\ldots,\tilde{j}_{k+1}; g) \\
&& \times \sum_{l_{k+2},\ldots,l_{2k+2}}' f(\tilde{j}_{k+2},\ldots,\tilde{j}_{2k+2}; g) \},
\end{eqnarray}
where $\tilde{j}_i = j_i + l_i N_\phi$. Thus, the factorization is not quite complete in this case because the sums over $\{ l_{i\leq k+1} \}$ and $\{ l_{i>k+1} \}$ remain coupled via the constraint that both of them are ranging only over integers $\equiv g \; {\rm mod} \; (k+1)$. This is a direct consequence of the periodic boundary condition Umklapp processes. Such an expression nevertheless allows for a dramatic reduction in computation time in diagonalizing these Hamiltonians, especially for $n>2$-body interactions, and provides insight into their analytic structure, as we explain in Sec.~\ref{sec:rr}.

\subsection{Outline of the approach and notations}\label{sec:approach}
 
Before analyzing concrete examples in Secs.~\ref{sec:rr},\ref{sec:nonun},\ref{sec:spin}, we would like to summarize the general approach and our notational conventions. We will be considering different families of Hamiltonians, like those in Eq.~(\ref{RRHamiltonian}),  expressed in the second-quantized form (an example for the two-body case is given in Eq.~(\ref{2bodygen})). The second-quantized Hamiltonians are written in terms of operators $c_j^\dagger$, which create a particle in the state $|j\rangle$, where $j$ is an integer ranging over the available number of orbitals. 

Hamiltonians projected to a Landau level possess a general symmetry of momentum conservation: the process of scattering between particles with indices $\{ j_i \}$ into those with indices $\{ j'_i \}$ is allowed only if $\sum_i j_i = \sum_{i} j'_i$. The equality is exact for a cylinder, and valid up to modulo $N_\phi$ on a torus. It is useful to introduce a number
\begin{equation}
p=\frac{\sum_{i=1}^{k+1} j_i}{k+1} \in \mathbb{Z}/(k+1)
\end{equation}
which labels the center of mass of a $k+1$-particle cluster, which is conserved up to a possible modulo $N_\phi$. Therefore, $j_i=p+j_i^{\rm rel}$, and $j_i^{\rm rel}$ must be an integer divided by $k+1$. Because we consider translationally-invariant interactions, their matrix elements do not depend on $p$ but only on $j_i^{\rm rel}$, which are of the form $\mathbb{Z}/(k+1)$. For 2-body interactions, $j_i^{\rm rel}$ are integers or half-integers (i.e. $p$ can be one of the orbitals or exactly half-way between two neighboring orbitals), for 3-body interactions we get integers or fractions with denominator 3, etc. We adopt this unusual choice of labelling because it allows one to immediately read off the value of the interaction matrix element for each type of scattering processes.

After obtaining the second-quantized form of the Hamiltonian, it becomes possible to perform an expansion in terms of $\kappa$:
\begin{equation}\label{expansion}
H=\sum_m \mathcal{P}(\kappa) e^{-\kappa^2 m^2} \sum_{i_1<i_2<...<i_{k+1}}\hat{P}^m_{i_1,...,i_{k+1}},
\end{equation}  
where $\mathcal{P}$ is at most a polynomial in $\kappa$ (i.e., contains no exponential factors in $\kappa$), and $\hat P$ is an operator containing $k+1$ creation and annihilation terms. Operator $\hat P$ contains information about the geometry of the manifold, while the prefactor depends on the specific form of the interaction.

A minimal number of terms in the expansion (\ref{expansion}) that is required to recover a complete set of zero-energy thin-torus patterns for a given state is referred to as the ``minimal truncated Hamiltonian":
\begin{equation}\label{expansion2}
H'=\sum_m^\Lambda \mathcal{P}(\kappa) e^{-\kappa^2 m^2} \sum_{i_1<i_2<...<i_{k+1}}\hat{P}^m_{i_1,...,i_{k+1}},
\end{equation}
The value of $\Lambda$ defining the minimal Hamiltonians for the Read-Rezayi $\mathbb{Z}_k$ states is such that $H'$ gives rise to $k+1$ degenerate ground states, which coincides with the well-known ``thin-torus limit" in the literature~\cite{Bergholtz-PhysRevB.77.155308,Bergholtz-PhysRevLett.99.256803,Seidel-PhysRevLett.95.266405}. The ground state, as well as the excited states, of such Hamiltonians are also the exact eigenstates of the full Hamiltonian $H$ in the limit $L_y \to 0$. In this sense, $H'$ ``approximates" the full Hamiltonian $H$ when $L_y$ is vanishingly small.

However, corrections beyond this trivial limit can also be obtained by setting somewhat larger values for $\Lambda$. As we illustrate below, such corrections can generally be organized in a positive semi-definite form:
\begin{equation}\label{posdef}
H'(\Lambda) = \sum_p A_p^\dagger A_p,
\end{equation}
so that the (truncated) ground-state energy $E'(\Lambda) \geq 0$. For certain choices of $\Lambda$, though not generally, the ground-states $\Psi'(\Lambda)$ of $H'$ can be analytically computed, and turn out to be exact zero modes of $H'$, \emph{as well as} $H$. Such states can be viewed as approximations to the true ground state of $H$ -- their overlap with the true ground state of $H$ typically increases monotonically as $\Lambda$ is increased. A relatively small value of $\Lambda$ is empirically found to be sufficient to obtain extremely accurate approximations to the ground state even at substantial values of $L_y$ (see, e.g., Sec.~\ref{sec:factorization}). 
However, for larger values of $\Lambda$, the solutions $\Psi'(\Lambda)$ are unlikely to have zero energy, and obtaining their analytic form appears more difficult and may necessitate the use of perturbation theory (or degenerate perturbation theory for the excited states).

In the remainder of this article, we analytically solve for the eigenenergies and eigenstates of $H'$ in several tractable cases. The obtained solutions for energies and eigenfunctions are tested against numerical solutions of the \emph{full} Hamiltonians in finite systems and small $L_y$ regime. We generally find a range of $L_y$ where the truncated Hamiltonians capture accurately the physics of the system described by the full Hamiltonian. To avoid any confusion, we emphasize that the energy spectra shown in Figures below are always computed by numerical (exact) diagonalization of small finite systems, while the analytic solutions discussed in the text are valid for any system size. 

Finally, we note that in addition to the thin torus, we will also consider thin cylinders where FQH states typically have a unique ground state (which simplifies the analysis). The expansion of the Hamiltonian (\ref{expansion}) is formally similar in both cases (matrix elements for thin torus and thin cylinder are nearly identical in value because the interaction of a particle with its mirror images is strongly suppressed), however the torus Hamiltonian contains explicit terms where the particle at site $N_\phi-1$ interacts with particle at site $0$, etc.

\section{Read-Rezayi series}\label{sec:rr}

The bosonic Hamiltonian of Eq.~(\ref{RRHamiltonian}) that describes the Read-Rezayi states, including the Laughlin and Moore-Read Pfaffian state, has the following matrix element on a cylinder:
\begin{eqnarray}
V_{j_1,\ldots,j_{2k+2}} &=& 
%\exp\Big\{-\frac{1}{2} \left[ \sum_i X{j_i}^2 - \frac{1}{2k}( \sum_i X{j_i} )^2 \right] \Big\}\nonumber\\
\exp\Big\{-\frac{\kappa^2}{2} \left[ \sum_i j_i^2 - \frac{(\sum_i j_i )^2}{2k+2} \right] \Big\}.\label{matel}
\end{eqnarray} 
We omitted the normalization which is defined by requiring that any $k+1$-particle droplet has energies 0 or 1 in the thermodynamic limit. Matrix element for the torus geometry can be obtained following the steps outlined in Appendix~\ref{sec_app}. By transforming to the relative and center-of-mass coordinate frame,  
Eq.~\ref{matel} can be decoupled in the form Eq.~\ref{torusmatel}. Then, by direct inspection of Eq.~\ref{matel} and counting the powers in the exponent, we can perform the expansion around the thin-cylinder limit, such as in Eq.~(\ref{expansion}),
i.e., identify the dominant terms as $\kappa \to \infty$. 

\subsection{Laughlin state}\label{sec:laughlin}

For the bosonic Laughlin state at $\nu=1/2$, the leading scattering processses (in the order of decreasing amplitude) are:
\begin{eqnarray}\label{laughlin_trunc}
\nonumber && {c_p^\dagger}^2 c_p^2; \; \sim 1; \; \cancel{2}, \\
\nonumber && c_{p+\frac{1}{2}}^\dagger c_{p-\frac{1}{2}}^\dagger c_{p-\frac{1}{2}} c_{p+\frac{1}{2}}; \; \sim e^{-\kappa^2/2}; \; \cancel{11}, \\
\nonumber && c_{p+1}^\dagger c_{p-1}^\dagger c_p^2; \; \sim e^{-\kappa^2}; \; 020\leftrightarrow 101, \\
\nonumber && c_{p+1}^\dagger c_{p-1}^\dagger c_{p-1} c_{p+1}; \; \sim e^{-2\kappa^2}; \; \cancel{101}, \\
&& \ldots
\end{eqnarray}
As we mentioned in Sec.~\ref{sec:approach}, operator $c_\alpha^\dagger$ creates an electron in the single-particle state $\alpha$,  thus in the present case $p$ must be an integer or half-integer. The order of magnitude of each type of interaction process in indicated next to each term in Eq.~\ref{laughlin_trunc}. 

Throughout this paper, we use the following notation. The crossed-out symbols, such as $\cancel{2}$, pictorially represent density-density type terms ${c_p^\dagger}^2 c_p^2$ that prevent (give energy to) the appearance of a certain pattern ($2$ in the present case) at any location $p$. Of course, these terms (for bosons) also imply that energy penalty will be incurred for creating configurations 3, 4, etc. particles in the same orbital. The many-body pair-hopping terms are depicted by arrows, e.g. $020\leftrightarrow 101$, and always imply the hermitian conjugates as well,  $101\leftrightarrow 020$. As with the density-density terms, the notation represents a \emph{minimal} process that can take place, but other allowed processes such as $112\leftrightarrow 031$ are also implied. Observe that for large $\kappa$ all the terms in Eq.~(\ref{laughlin_trunc}) are separated in a  hierarchy of energy scales $\exp(-\alpha\kappa^2)$.

\begin{figure}[htb]
\centerline{\includegraphics[width=\linewidth]{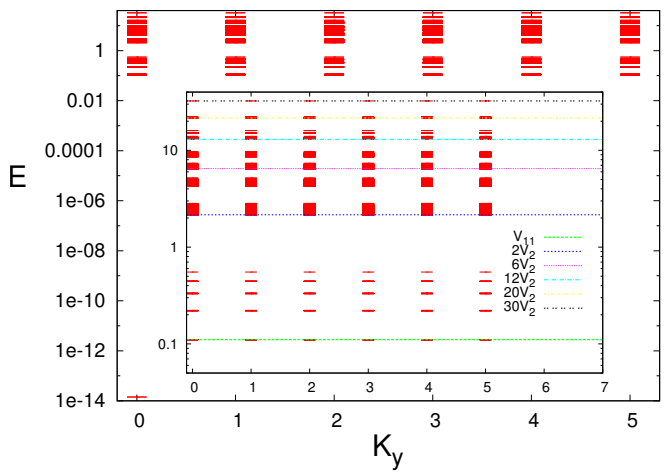}}
\caption[]{(Color online). Energy spectrum (log scale) of the full Laughlin Hamiltonian for 6 bosons and 12 flux quanta, torus aspect ratio 1/14. Inset: zoom on the spectrum above the ground state. Lines indicate the values of the corresponding density-density matrix elements of the Hamiltonian, which define the classical estimates for the energies of the excited states of the truncated model~\ref{laughlin_trunc}. These estimates show excellent agreement with the exact energies of the full Hamiltonian. For the purpose of clarity, spectrum is only plotted versus $K_y$ quantum number.}
\label{fig_laughlin}
\end{figure}

As previously pointed out~\cite{Jansen-2012JMP....53l3306J,Soule-PhysRevB.85.155116,Nakamura-PhysRevLett.109.016401,Wang-PhysRevB.87.245119}, keeping the first two types of terms in Eq.~(\ref{laughlin_trunc}) gives rise to a zero-energy ground state (twofold degenerate on the torus), which is the single permanent (i.e., bosonic occupation state) $101010\ldots$. Slightly away from the thin-torus limit, this configuration evolves into the one dressed by ``squeezing"~\cite{Nakamura-PhysRevLett.109.016401}, still at zero energy:
\begin{eqnarray}\label{laughlings}
\Psi_0 = \prod_p \{ 1 - \sqrt{2} e^{-\kappa^2} {c_p^\dagger}^2 c_{p+1}c_{p-1}\} |1010\ldots\rangle.
\end{eqnarray}
That this is also a zero-energy ground-state of the truncated Hamiltonian, Eq.~(\ref{laughlin_trunc}), can be proved by noting that the truncated Hamiltonian can be expressed in the positive semidefinite form 
\begin{eqnarray}\label{semidefinite}
&& H = \sum_{p \in \mathbb{Z}} A_p^\dagger A_p + \sum_{p \in \mathbb{Z}+\frac{1}{2}} B_p^\dagger B_p, \\
&& A_p^\dagger={c_p^\dagger}^2+2\exp(-\kappa^2)c_{p+1}^\dagger c_{p-1}^\dagger, \\
&& B_p^\dagger=2\exp(-\kappa^2/4)c_{p+1/2}^\dagger c_{p-1/2}^\dagger,
\end{eqnarray}
such that $A_p\Psi_0 = B_p \Psi_0=0$ for all $p$. Multiplicative factors of 2 and 4 in this equation come from the bosonic commutation relations. 

State $\Psi_0$ in Eq.~(\ref{laughlings}) has excellent overlap in finite-size systems with the ground-state of the full Hamiltonian when $\kappa$ is large. The intuition behind the statement that $\Psi_0$ remains a zero-energy ground state in the presence of hopping is the following. $\Psi_0$ contains a configuration $1010101010...$, as well as all the ones where locally $101$ has hopped to $020$. The latter configurations violate the density-density repulsion term $\cancel{2}$, therefore one would expect they incur an energy cost $\sim 1$. However, now that the state contains  both $101$ and $020$ droplets, the hopping process $101\leftrightarrow 020$ is active, and can lower the energy. Because the Hamiltonian is tuned in a fine way such that the magnitude of the hopping $t=e^{-\kappa^2}$ is exactly equal to the square root of the product of two density-density type terms, $\cancel{2}$ ($V_{\text{\cancel{2}}}=1$) and $\cancel{101}$ ($V_{\text{\cancel{101}}}= e^{-2\kappa^2}$), the energy of such configurations can be brought back to zero. In other words, if we look at this problem as a two-level system, the condition
\begin{equation}
\det \left(\begin{array}{cc}
V_{\text{\cancel{2}}} & t \\
t & V_{\text{\cancel{101}}} \\
\end{array} \right) = 0,\label{eq:zeromodecondition}
\end{equation} 
gives a null mode. This subtle factorization property is responsible for being able to express the Hamiltonian in positive semidefinite form (\ref{semidefinite}), and holds for the matrix elements of any Read-Rezayi state but not in general for other states like the Gaffnian or Haffnian. 

In Refs.~\cite{Nakamura-PhysRevLett.109.016401,Bergholtz2011755,Wang-PhysRevB.86.155104}, the solvable model defined by Eq.~(\ref{laughlin_trunc}) was used to construct an effective spin-1 chain for the Laughlin state by mapping $02 \rightarrow |1\rangle$, $10 \rightarrow |0\rangle$, and $00 \rightarrow |-1\rangle$. From this mapping, the ground state Eq.~(\ref{laughlings}) was rewritten as a matrix-product state (MPS)~\cite{Jansen-2012JMP....53l3306J,Soule-PhysRevB.85.155116,Nakamura-PhysRevLett.109.016401,Wang-PhysRevB.87.245119}. This MPS, however, is different from one based on conformal field theory, and does not describe the system accurately in the large $L_y$ limit~\cite{Estienne-PhysRevB.87.161112}.

In contrast to previous works that primarily addressed the nature of the ground state, we find that the entire neutral energy spectrum in the thin-torus limit also has a simple form and splits into bands that can be classified according to the violation of $(k,r)=(1,2)$ clustering conditions, Fig.~\ref{fig_laughlin}. The ground-state is unique (up to the center-of-mass degeneracy), and  satisfies (1,2) clustering property -- no more than a single particle in each two consecutive orbitals. We can predict that the first group of excited states will have energies proportional to $\sim V_{\text{\cancel{11}}}$, i.e. it will contain states that locally contain $...11...$ patterns. More precisely, we obtain $N-1$ of such bands in the lowest part of spectrum, i.e. bands with energies $V_{\text{\cancel{11}}},...,(N-1)V_{\text{\cancel{11}}}$, as seen in the inset of Fig.~\ref{fig_laughlin}. The states in these bands still satisfy (2,2) clustering. 

The above band of excited states terminates when the pattern ...2... starts to appear, i.e., (2,1) clustering sets in. These states violate the first term in Eq.~(\ref{laughlin_trunc}), and therefore will have energies $\sim V_{\text{\cancel{2}}}$. We can predict that the energy of this band, relative to the one below it which satisfied (2,2) clustering, will be given by
\begin{equation}
\frac{E_{(2,1)}}{E_{(2,2)}} = \frac{2V_{\text{\cancel{2}}}}{4V_{\text{\cancel{11}}}}=\frac{1}{2}\exp(\kappa^2/2),
\end{equation}
which agrees very accurately with the exact-diagonalization result. Here the factors of 2 and 4 come from the action of the bosonic creation/annihilation operators and from their commutation relations. As shown in the inset of Fig.~\ref{fig_laughlin}, we again obtain several groups of these states that contain a number of ...2... patterns. This scenario continues with bands of states appearing that satisfy (3,1),(4,1),(5,1) and so on, eventually terminating with a single state ($N$,1). 

\begin{figure}[htb]
\centerline{\includegraphics[width=\linewidth]{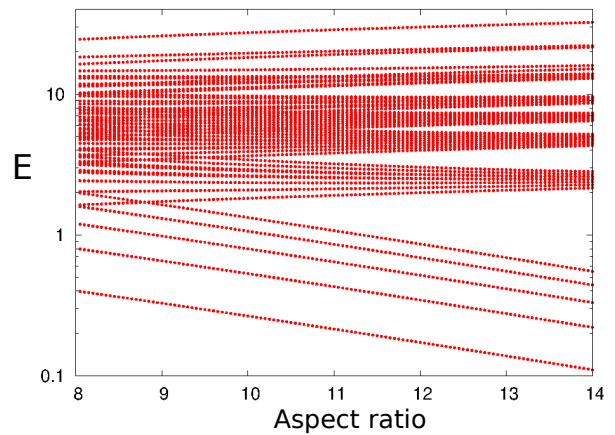}}
\caption[]{(Color online). Evolution of thin-torus energy bands (log scale) as a function of the aspect ratio, going from thin to isotropic limit. The bands remain stable for ratio larger than $\approx 12$.}
\label{fig_laughlin_ev}
\end{figure}
As we evolve the system towards the isotropic limit by changing the aspect ratio (Fig.~\ref{fig_laughlin_ev}), we find that the bands mentioned above remain stable down to aspect ratio $\approx 12$. For aspect ratios larger than this value, the dressed configurations of the lowest excited states are also constructed using the same ``squeezing" operator as in Eq.~(\ref{laughlings}), but choosing a different root configuration, e.g. 110101...0100 for one of the first excited states and so on. To describe aspect ratios smaller than $\approx 12$, Eq.~(\ref{laughlin_trunc}) is no longer sufficient, and we must keep additional terms in the expansion. However, in this case a question immediately presents itself: if we keep additional terms in the expansion, can the new truncated Hamiltonian also be written in a positive semidefinite form? 

This question is analyzed in detail in the Appendix~\ref{sec:factorization}. It is shown that truncating the Hamiltonian at some order in $\kappa$, in general, does not allow one to exactly rewrite it as $\sum_p A_p^\dagger A_p$, except at a very low order of the truncation. 

However, instead of expanding $H$, one can directly expand the $A_p^\dagger$ operator, like it is done in Eq.~\ref{laughlin_fact}. By expanding $A_p^\dagger$ to the order $\exp(-\alpha \kappa^2)$, we generate a positive semidefinite Hamiltonian $\tilde H = \sum_p A_p^\dagger A_p$ that ``approximates" the full Hamiltonian $H$ to the order $\exp(-2\alpha\kappa^2)$, in the sense that the eigenstates of $\tilde H$ have large overlap with those of $H$. This can be verified numerically (see Fig.~\ref{fig_laughlin_truncspectrum}). In this way, we can generate a family of positive semidefinite Hamiltonians $\tilde H$, whose eigenstates monotonically approach those of the full Hamiltonian $H$. However, the ground-state energy of the truncated Hamiltonian $\tilde H$ is not guaranteed to be strictly zero and varies non-monotonically as a function of $\kappa$.
 
\subsection{Moore-Read state}\label{sec:mr}

The solvable Hamiltonian for the bosonic Moore-Read state contains the following terms, listed in the order of dominance:
\begin{eqnarray}\label{mr_trunc}
\nonumber && {c_p^\dagger}^3 c_p^3; \; \sim 1; \; \cancel{3}, \\
\nonumber && c_{p+\frac{2}{3}}^\dagger {c_{p-\frac{1}{3}}^\dagger}^2 {c_{p-\frac{1}{3}}}^2 c_{p+\frac{2}{3}}; \; \sim e^{-2\kappa^2/3}; \; \cancel{21}, \\
\nonumber && c_{p+1}^\dagger c_p^\dagger c_{p-1}^\dagger c_p^3; \; \sim e^{-\kappa^2}; \; 030\leftrightarrow 111, \\
 \nonumber && {c_{p+\frac{2}{3}}^\dagger}^2 c_{p-\frac{4}{3}}^\dagger {c_{p-\frac{1}{3}}}^2 c_{p+\frac{2}{3}}; \; \sim e^{-5\kappa^2/3}; \; 021\leftrightarrow 102, \\
\nonumber && c_{p+1}^\dagger c_p^\dagger c_{p-1}^\dagger c_{p-1} c_p c_{p+1}; \; \sim e^{-2\kappa^2}; \; \cancel{111}, \\
\nonumber && {c_{p+\frac{2}{3}}^\dagger}^2 c_{p-\frac{4}{3}}^\dagger c_{p-\frac{4}{3}} {c_{p+\frac{2}{3}}}^2; \; \sim e^{-8\kappa^2/3}; \; \cancel{102}, \\
&& \ldots
\end{eqnarray}
This Hamiltonian is positive semidefinite and coincides with the full Moore-Read Hamiltonian at orders above $\exp(-8\kappa^2/3)$; exactly at this order, it misses a single term given in Eq.~(\ref{neglect}). More details on the derivation of the above Hamiltonian and its positive semidefinite property are given in Appendix~\ref{sec:factorization}. We proceed to solve for the spectrum of the Hamiltonian (\ref{mr_trunc}) in the vicinity of the thin-torus limit. Appendix~\ref{sec:factorization} contains numerical evidence that the solvable Hamiltonian in Eq.~\ref{mr_trunc} gives an accurate description of the full problem near the thin-torus limit.

As we emphasized above, the minimal truncated Hamiltonian for the Moore-Read state contains the first two types of terms in Eq.~(\ref{mr_trunc}). They prevent patterns ...3... and ...21..., therefore the ground state is such that each two consecutive orbitals can have at most two particles. The exact zero modes are then 202020..., 020202..., and 11111..., which represent the three-fold degenerate manifold of the Moore-Read ground-state that satisfies (2,2) clustering property. 

\begin{figure}[t]
\centerline{\includegraphics[angle=270,width=\linewidth]{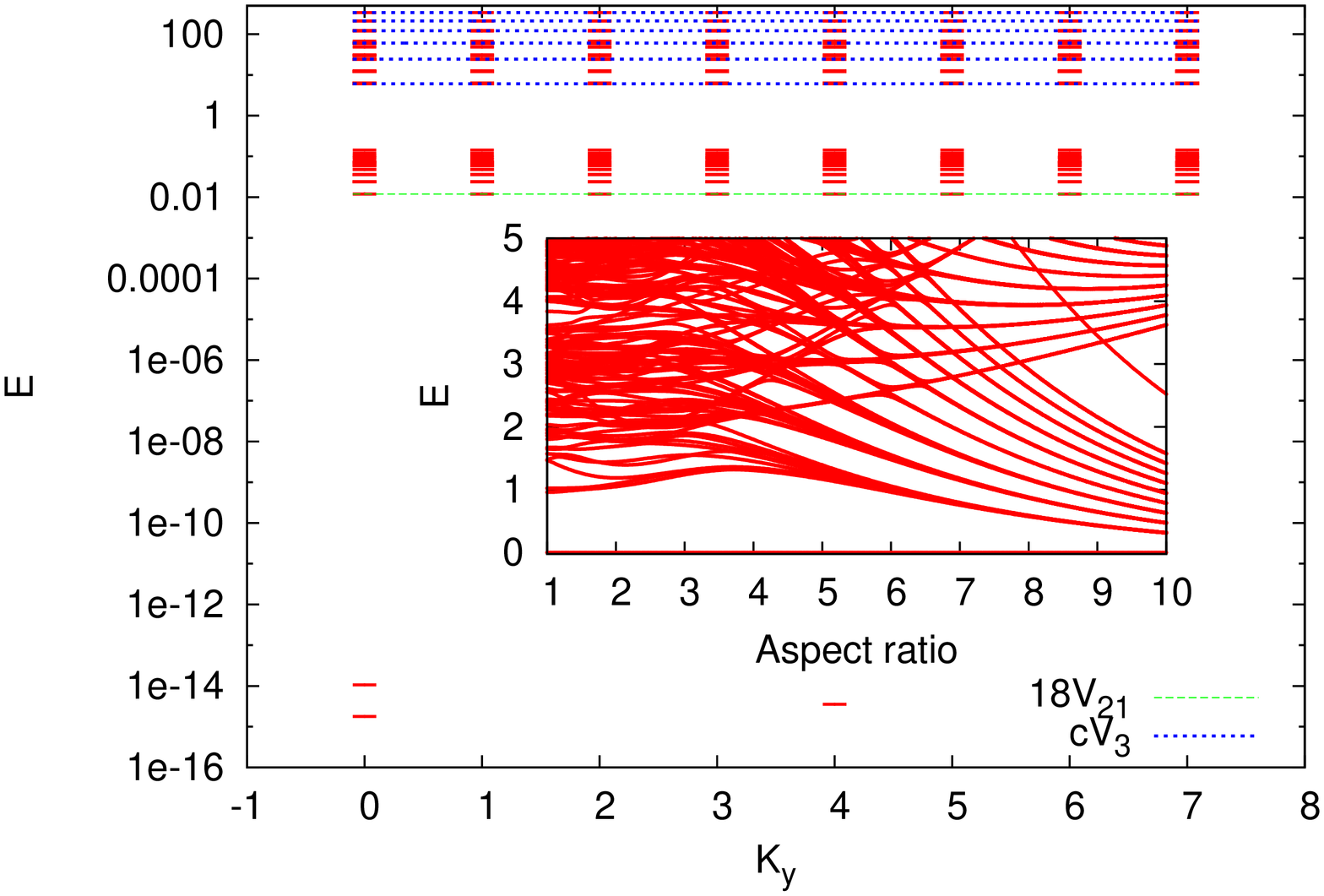}}
\caption[]{(Color online). Energy spectrum (log scale) of the full Moore-Read Hamiltonian for 8 bosons and 8 flux quanta, torus aspect ratio 1/14. Lines indicate the values of the corresponding density-density matrix elements of the Hamiltonian,  $18V_{\text{\cancel{21}}}$ and multiples of $6V_{\text{\cancel{3}}}$, which define the classical estimates for the energies of the excited states of the truncated model~\ref{mr_trunc}. These estimates show excellent agreement with the exact energies of the full Hamiltonian. The three low-lying states are the topologically degenerate Moore-Read ground states (they appear to be split because of the numerical precision limitations). For the purpose of clarity, spectrum is only plotted versus $K_y$ quantum number. Inset: energy spectrum as a function of the aspect ratio showing the transition from the thin torus limit to the isotropic 2D limit. }
\label{fig_mr}
\end{figure}
Because the minimal truncated Hamiltonian is effectively classical, we can again solve for the energies of all states in the spectrum with excellent agreement with the numerical result for the full Moore-Read Hamiltonian, Fig.~\ref{fig_mr}. The first excited state of the Moore-Read Hamiltonian is separated from the ground state manifold by a gap
\begin{equation}\label{mr_gap}
E_{(3,2)}=18\exp(-2\kappa^2/3),
\end{equation}
proportional to the amplitude of the term $V_{\text{\cancel{21}}}$ (with the prefactor 18, resulting from bosonic statistics). This is indeed found in the exact spectrum, Fig.~\ref{fig_mr}. The first excited state belongs to a band of states that satisfy (3,2) clustering. The next band of states obeys (4,2) clustering and so on. Eventually, (3,1) states start to appear at a new energy scale which is given by $V_{\text{\cancel{3}}}$. More precisely, the energy of this band of states, relative to those that satisfy (3,2) clustering, can be estimated to be given by
\begin{equation}
\frac{E_{(3,1)}}{E_{(3,2)}} = \frac{6V_{\text{\cancel{3}}}}{18V_{\text{\cancel{21}}}}=\frac{1}{3}\exp(2\kappa^2/3),
\end{equation}
again in excellent agreement with the numerical data. Depending on $N$, multiple (3,1) clusters are allowed, until (4,1) states start to appear, etc. The spectrum terminates with a single state that obeys ($N$,1) clustering.

In order to generate the first-order correction to the thin-torus limit, we must keep all the terms in the Hamiltonian listed in Eq.~(\ref{mr_trunc}). This Hamiltonian can also be written in a positive semidefinite form:
\begin{eqnarray}\label{semidefinitemr}
H &=& \sum_{p \in \mathbb{Z}} A_p^\dagger A_p + \sum_{p \in \mathbb{Z}+\frac{1}{3}} B_p^\dagger B_p, \\
A_p^\dagger &=& {c_p^\dagger}^3 + 6\exp(-\kappa^2)c_{p+1}^\dagger c_p^\dagger c_{p-1}^\dagger, \\
\nonumber  B_p^\dagger &=& 3\exp(-\kappa^2/3) c_{p+2/3}^\dagger {c_{p-1/3}^\dagger}^2 \\ 
 && + 3\exp(-4\kappa^2/3) {c_{p+2/3}^\dagger}^2 {c_{p-4/3}^\dagger}.
\end{eqnarray}
From this, we can infer the dressed solutions, similar to Eq.~(\ref{laughlings}):
\begin{eqnarray}\label{mrgs}
&& \prod_p \{ 1 - \sqrt{2} e^{-\kappa^2} {c_{p-\frac{1}{3}}^\dagger}^2 c_{p+\frac{2}{3}}^\dagger {c_{p+\frac{2}{3}}}^2 c_{p-\frac{4}{3}}  \} |2020\ldots\rangle, \\
&& \prod_p \{ 1 - \sqrt{6} e^{-\kappa^2} {c_{p}^\dagger}^3 c_{p+1} c_{p} c_{p-1}  \} |1111\ldots\rangle
\end{eqnarray}
These states are annihilated by $A_p$ and $B_p$ for all $p$. Alternatively, we can show they are zero modes by considering an equivalent two-level system, analogous to Eq.~(\ref{eq:zeromodecondition}) (now there are two such systems, because different ground states ``live" in different momentum sectors). For one type of the ground states, $t^2\equiv V_{030\leftrightarrow 111}^2 = V_{\text{\cancel{3}}} V_{\text{\cancel{111}}}$, and in the other sector $t^2\equiv V_{021\leftrightarrow 102}^2 = V_{\text{\cancel{21}}} V_{\text{\cancel{102}}}$. Because of this factorization property, the dressed wave functions are solutions of zero-energy, similar to the Laughlin case.  

\subsection{Read-Rezayi states}\label{sec:z3}

Read-Rezayi $\mathbb{Z}_3$ Hamiltonian expanded at the order $\exp(-\kappa^2)$ is given by the following terms
\begin{eqnarray}\label{z3_trunc}
\nonumber && {c_p^\dagger}^4 c_p^4; \; \sim 1; \; \cancel{4}, \\
\nonumber && c_{p-\frac{3}{4}}^\dagger {c_{p+\frac{1}{4}}^\dagger}^3 {c_{p+\frac{1}{4}}}^3 c_{p-\frac{3}{4}}; \; \sim e^{-3\kappa^2/4}; \; \cancel{31}, \\
 && {c_{p+\frac{1}{2}}^\dagger}^2 {c_{p-\frac{1}{2}}^\dagger}^2 c_{p-\frac{1}{2}}^2 c_{p+\frac{1}{2}}^2; \; \sim e^{-\kappa^2}; \; \cancel{22}, \\
\nonumber && c_{p+1}^\dagger {c_p^\dagger}^2 c_{p-1}^\dagger {c_{p}}^4, \; \sim e^{-\kappa^2}; \; 040 \leftrightarrow 121, \\
\nonumber&& \ldots
\end{eqnarray} 
The first three density-density terms yield the well-known degenerate Read-Rezayi ground-states, 3030... and 2121... (and their center-of-mass copies), that satisfy (3,2) clustering. Notice, however, that there is a hopping term of exactly the magnitude $\exp(-\kappa^2)$, and thus must be kept along with the density-density terms. The presence of the hopping term complicates the problem because the Hamiltonian (\ref{z3_trunc}) is no longer expressible in a positive semidefinite form. One might hope that, by keeping more terms in the Hamiltonian expansion, it would be possible to obtain a closed form of the truncated Hamiltonian, but one can show this does not happen by following the argument outlined in Appendix~\ref{sec:factorization}.

The RR case is thus different from the Laughlin and Moore-Read cases because the ``minimal" Hamiltonian that describes the thin-torus limit is not ``protected" from the hopping terms, i.e., it becomes intrinsically \emph{non-classical} and its properties, such as the existence of a gap, become less obvious. Given that the hopping term, Eq.~(\ref{z3_trunc}), is relatively large in magnitude (comparable to one of the density-density terms), the first question we would like to address is whether the full RR Hamiltonian, truncated at the order $\exp(-\kappa^2)$, might become gapless in the presence of this hopping.

\begin{figure}[htb]
\centerline{\includegraphics[angle=270,width=\linewidth]{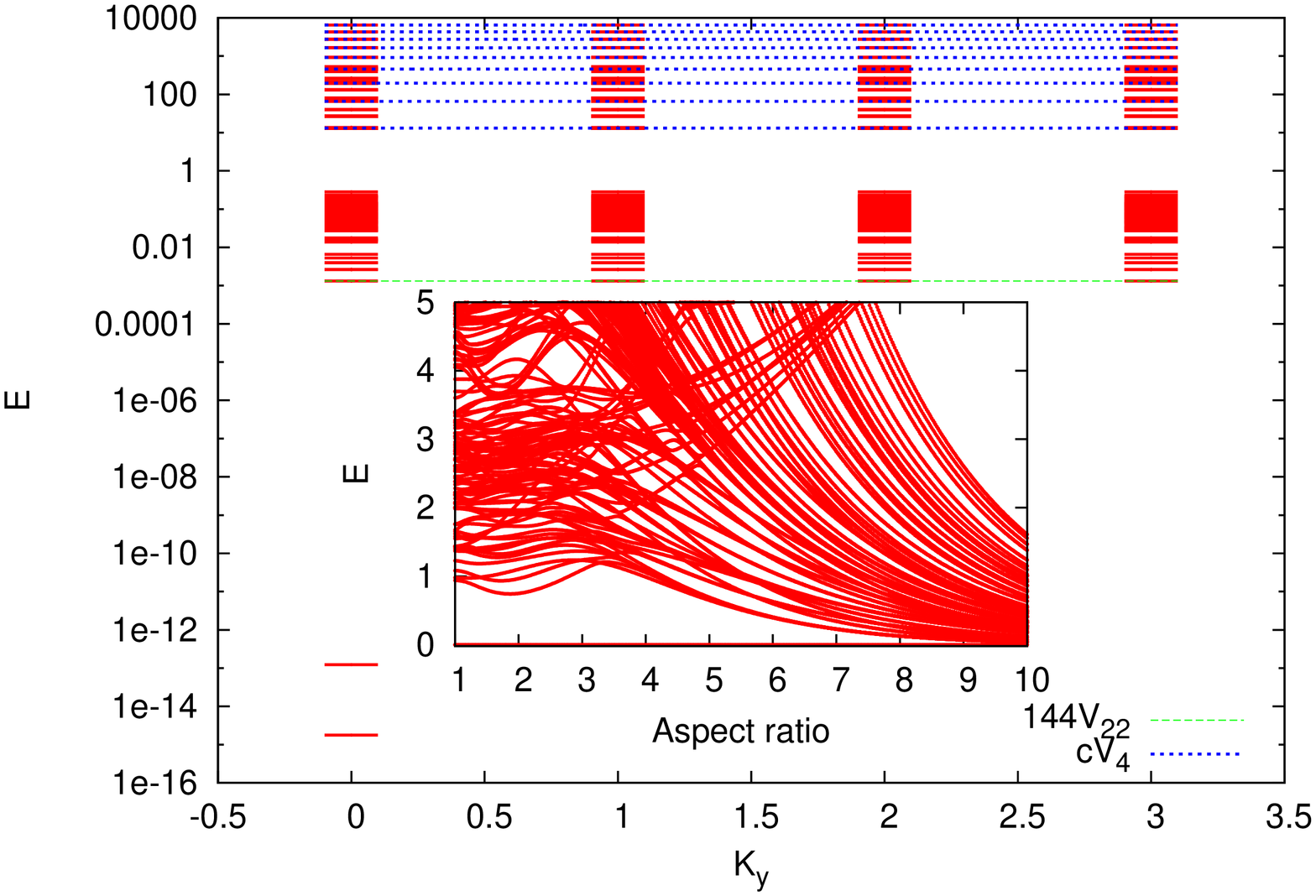}}
\caption[]{(Color online). Energy spectrum (log scale) of the full $\mathbb{Z}_3$ Read-Rezayi Hamiltonian for 12 bosons and 8 flux quanta, torus aspect ratio 1/14. Lines indicate the values of the corresponding density-density matrix elements of the Hamiltonian,  $144V_{\text{\cancel{22}}}$ and multiples of $24V_{\text{\cancel{4}}}$, which define the classical estimates for the energies of the excited states of the truncated model~\ref{z3_trunc}. The two low-lying states are the topologically degenerate Read-Rezayi ground states (they appear to be split because of the numerical precision limitations). For the purpose of clarity, spectrum is only plotted versus $K_y$ quantum number. Inset: energy spectrum as a function of the aspect ratio showing the transition from the thin torus limit to the isotropic 2D limit.}
\label{fig_rr}
\end{figure}

Similarly to the Laughlin and Moore-Read cases, the first excited state of the RR Hamiltonian (the quasiparticle-quasihole pair) is given by the pattern 12212121... (moving a single particle in the ground state pattern 21212121... to an adjacent orbital). This configuration contains a number of 121 droplets that might hop to 040. However, the maximum energy reduction due to such a process is on the order of $V_{\text{\cancel{121}}}$ density-density term, which can be shown to be $\exp(-2\kappa^2)$. Since the configuration also contains 22 terms, which come with an energy penalty $\exp(-\kappa^2)$,
the energy gap of the truncated RR Hamiltonian is given by
\begin{equation}\label{rr_gap}
E_{(4,2)}=144 \exp(-\kappa^2) - O(\exp(-2\kappa^2)),
\end{equation}
and the hopping $121\leftrightarrow 040$ is not able to close the gap. Exact diagonalization calculations confirm that the system is always gapped, and the value of the gap matches $\exp(-\kappa^2)$, Fig.~\ref{fig_rr}. 

The particular example of $\mathbb{Z}_3$ state illustrates that generally, as we go towards higher-$k$ Read-Rezayi states, numerous hoppings begin to enter even the minimal truncated Hamiltonian (the Hamiltonian with the smallest number of density-density terms required to obtain the thin-torus root states). Thus, one might suspect that above some critical $k$, the Read-Rezayi states will completely cease to behave classically in the \emph{thin-torus limit itself}, which could be a manifestation of their defficiency to screen, etc. 

However, such ``dangerous" hoppings emerge rather gradually. For example, states up to $k=6$ only have a single hopping, $0k0\leftrightarrow 1(k-2)1$ in the minimal truncated Hamiltonian. In the case of $\mathbb{Z}_{k\geq 7}$, we find more than one hopping; however, for the same reason as above, none of these hoppings can close the gap. Since the lowest excited state contains a droplet $0\frac{k+1}{2}\frac{k+1}{2}$ (for $k$-odd), one might envision a hopping $0\frac{k+1}{2}\frac{k+1}{2} \leftrightarrow 1 \frac{k-3}{2}\frac{k+3}{2}$, but it can be shown that the energy gain in that case is lower several orders of magnitude than the density-density term $\cancel{\frac{k+1}{2}\frac{k+1}{2}}$ and therefore only leads to the fine splittling of the thin-torus energy levels, but cannot close the gap.  

We have verified by exact diagonalization on small systems that all states up to $\mathbb{Z}_9$ are gapped in the sense described above. The value of the gap is equal to the density-density term $\cancel{\frac{k+1}{2}\frac{k+1}{2}}$ for $k$-odd, and $\cancel{(\frac{k}{2}+1)\frac{k}{2}}$ for $k$-even. Similar conclusion holds for the fermionic version of Read-Rezayi states. We emphasize that these results are for the systems near the thin-torus (1D) limit, and it is not obvious what they imply for the isotropic (2D) limit.

Before we proceed to analyze the states whose underlying CFT is not rational and unitary, we mention in passing that further approximations in $\kappa$ can be generated for the full RR Hamiltonian by keeping more terms in $A_p^\dagger$, Eq.~(\ref{rr_fact}), similarly to the Moore-Read state (Sec.~\ref{sec:mr}). 

\section{Non-unitary and irrational states}\label{sec:nonun}

\subsection{Gaffnian}

An intriguing non-unitary FQH state exists for $\nu=2/3$ filling of bosons under the name of Gaffnian~\cite{Simon-PhysRevB.75.075317}. This state in many ways behaves like a ``proper" FQH state (e.g., it even shares part of the entanglement spectrum with the Jain composite fermion state~\cite{Regnault-PhysRevLett.103.016801}, and is directly related to the unpolarized version of the $2/3$ hierarchy state~\cite{Milovanovic-PhysRevB.82.035316,Milovanovic-PhysRevB.80.155324}). However, elaborate arguments~\cite{Read-PhysRevB.79.245304,Toke-PhysRevB.80.205301,Freedman-PhysRevB.85.045414} have been put forward to show that this state cannot describe a gapped phase of matter. Here we would like to explore whether this has a more transparent manifestation near the thin-torus limit. 

Bosonic Gaffnian is defined by the 3-body clustered Hamiltonian~\cite{Simon-PhysRevB.75.075318} which is explicitly written out in Appendix~\ref{sec_app}. Each matrix element of the Gaffnian Hamiltonian contains a Gaussian term,  identical to that of the Moore-Read state, but in addition it has a quartic polynomial multiplying the Gaussian. However, near the thin-torus limit, the hiearchy of energy scales is controlled by the Gaussian only, and the dominant interaction terms of the Gaffnian Hamiltonian are in fact the same ones written in Eq.~(\ref{mr_trunc}) for the Moore-Read state. As we will see, because of a different filling factor, the resulting physics will also be very different from the previous cases. 

While for the Moore-Read case it was sufficient to keep only the first two terms in Eq.~(\ref{mr_trunc}) to reproduce the ground-state patterns, in the case of the Gaffnian we must keep all terms listed in Eq.~(\ref{mr_trunc}) to recover the correct set of Gaffnian ground states (200200..., 110110...). These ground states obey (3,2) clustering and are not affected by hoppings $030\leftrightarrow 111$ and $102\leftrightarrow 021$. 

Now, what is the first excited state $\Psi_1$ above the Gaffnian ground state, e.g. $\Psi_0=|\ldots 200200200 \ldots\rangle$? Naively, one would construct $\Psi_1$ by violating the fundamental (2,3) clustering condition. This can be done by nucleating one or more 201 droplets in the $\Psi_0$ pattern. The energy of such configurations would be proportional to $V_{\text{\cancel{102}}}$. As we will show, the true excited state indeed derives from a pattern such as $201100200200200\ldots$, however the classical estimate for the energy of this state is incorrect due to the quantum nature of the Gaffnian Hamiltonian near the thin-torus limit. 

\begin{figure}[t]
\centerline{\includegraphics[width=\linewidth]{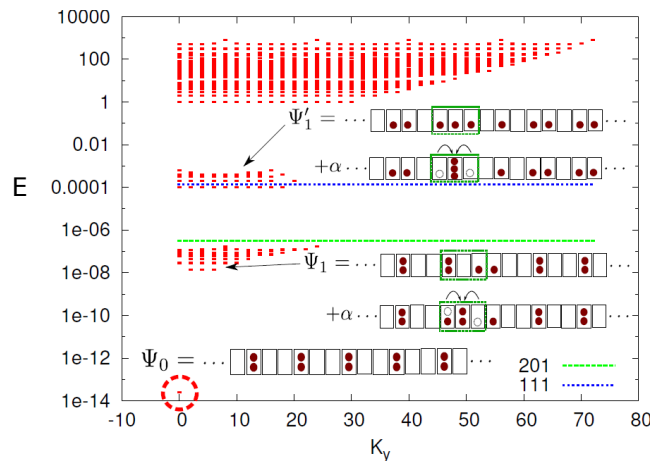}}
\caption[]{(Color online). Energy spectrum (log scale) of the full Gaffnian Hamiltonian for 8 bosons and 9 flux quanta, cylinder aspect ratio 0.07. Lines indicate the values of the corresponding density-density matrix elements of the Hamiltonian, which might be expected to define the energy bands of the truncated Hamiltonian (\ref{mr_trunc}). However, density-density terms $V_{\text{\cancel{102}}}$ and $V_{\text{\cancel{111}}}$ (denoted by lines) significantly overestimate the energy of the low-lying excited states $\Psi_1,\Psi_1'$, illustrating the importance of hopping terms played in the Gaffnian thin-torus description.}
\label{fig_gaff}
\end{figure}

It is instructive to consider a simple example of just 4 bosons on a cylinder. The ground-state root configuration in unique in that case and reads 2002. The sector of the Hilbert space with momentum 1 relative to the ground state contains only 4 states: $0130$, $0211$, $1021$ and $1102$. We can explicitly evaluate the Hamiltonian (\ref{mr_trunc}) in this basis
\begin{equation}\label{gaff_example}
\begin{bmatrix}
H_{11} & 6\sqrt{3}(3t+2t') & 0 & 0 \\
6\sqrt{3}(3t+2t') & H_{22} & 18t & 0 \\
0 & 18t & 18(V_{\text{\cancel{102}}}+V_{\text{\cancel{021}}}) & 18t \\
0 & 0 & 18t & 18 V_{\text{\cancel{102}}} 
\end{bmatrix}
\end{equation}
where $H_{11}=6(9V_{\text{\cancel{021}}}+V_{\text{\cancel{3}}})$, $H_{22}=18(V_{\text{\cancel{102}}}+4V_{\text{\cancel{111}}}+V_{\text{\cancel{021}}})$, $t,t'$ are the $021\leftrightarrow 102$ and $030\leftrightarrow 111$ hopping amplitudes, respectively. The lowest eigenstate can be accurately represented by considering the lower $2\times 2$-block of the Hamiltonian, corresponding to states $1102$ and $1021$ that form a two-level system. Considering the variational ansatz,
\begin{equation}
\psi= |1102\rangle - \alpha |1021\rangle,
\end{equation}
we find the expectation energy $\langle \psi|H|\psi \rangle/\langle \psi|\psi \rangle$
\begin{equation}
\frac{18 (2 \alpha t+ V_{\text{\cancel{102}}}+ \alpha^2 (V_{\text{\cancel{102}}}+V_{\text{\cancel{021}}}))}{1+\alpha^2},
\end{equation}
and the condition for this state to describe a gapless mode is
\begin{equation}
t^2-V_{\text{\cancel{102}}}^2-V_{\text{\cancel{102}}} V_{\text{\cancel{021}}}=0.
\end{equation}
This condition looks similar to the one for Laughlin and Moore-Read states; in the present case, however, it cannot be fulfilled because the matrix elements do not factorize:
\begin{eqnarray}
V_{\text{\cancel{021}}} &=& (3-4\kappa^2/3+4\kappa^4/9)\exp(-2\kappa^2/3), \\ 
V_{\text{\cancel{102}}} &=& (3-16\kappa^2/3+64\kappa^4/9)\exp(-8\kappa^2/3), \\ 
t &=& (3-10\kappa^2/3+16\kappa^4/9)\exp(-5\kappa^2/3), 
\end{eqnarray}
thus $t^2 \neq V_{\text{\cancel{021}}} V_{\text{\cancel{102}}}$. 

By directly minimizing the ground state energy, we find 
\begin{equation}
\alpha=\frac{V_{\text{\cancel{021}}} - \sqrt{4 t^2 + V_{\text{\cancel{021}}}^2}}{2 t},
\end{equation}
and the corresponding minimum
\begin{eqnarray}
\nonumber E_1 &=& 18 V_{\text{\cancel{102}}} + 9 V_{\text{\cancel{021}}} - 9 \sqrt{4t^2+V_{\text{\cancel{021}}}^2} \\
 && \approx 18 V_{\text{\cancel{102}}} - \frac{18 t^2}{V_{\text{\cancel{021}}}} + \frac{18t^4}{V_{\text{\cancel{021}}}^3} + \ldots
\end{eqnarray}
From this expression, we see that the classical energy
\begin{equation}
E_{\rm class} = 18 V_{\text{\cancel{102}}} = (128 \kappa^4-96 \kappa^2+54) \exp(-8\kappa^2/3),
\end{equation}
receives a quantum-mechanical correction due to hopping
\begin{eqnarray}
\nonumber \Delta E &\approx& - \frac{18 t^2}{V_{\text{\cancel{021}}}} + \frac{18t^4}{V_{\text{\cancel{021}}}^3} +\ldots \\
&&  \approx (-128 \kappa^4 + 96 \kappa^2+ 270 + \ldots) e^{-8\kappa^2/3} ,
\end{eqnarray}
which cancels some of the the polynomial terms in $\kappa$ in $E_{\rm class}$, but does not lead to an overall cancellation of terms. Therefore, the gap of the Gaffnian state is given by
\begin{equation}\label{gaff_gap}
E_1 = (324 + \frac{972}{\kappa^2} + O(\frac{1}{\kappa^4})) \exp(-8\kappa^2/3),
\end{equation}
where we neglected terms of the order $\exp(-14\kappa^2/3)$. We see that the gap is bounded by $e^{-8\kappa^2/3}$, though with a much smaller prefactor (a constant instead of $\kappa^4$) because of the hopping. Thus, we expect that the energies obtained by exact diagonalization of the Gaffnian Hamiltonian will show significant deviations from their classical predictions, but nevertheless remain bounded by $\exp(-8\kappa^2/3)$. This is indeed found in the (numerical) exact diagonalization of the full Gaffnian Hamiltonian near the thin-torus limit, Fig.~\ref{fig_gaff}.

In Fig.~\ref{fig_gaff} horizontal lines denote the values of $V_{\text{\cancel{102}}}$ and $V_{\text{\cancel{111}}}$ matrix elements, which represent the naive classical estimates for the energies of two groups of the excited states. One of the groups is the example considered above; the second group of states violates the condition of putting 3 particles in 3 consecutive orbitals. The classical energies overestimate the true energies of the excited states of Eq.~(\ref{gaff_excited}), schematically labeled as $\Psi_1,\Psi_1'$; the energy difference is given by the hopping term contributions $-t^2/V_{\text{\cancel{021}}}$ and $-t'^2/V_{\text{\cancel{3}}}$. A general expression for the wave functions of these two types of excitations of the Gaffnian in the thin torus limit is 
\begin{eqnarray}\label{gaff_excited}
\nonumber \prod_p \{1 - \alpha {c_{p-\frac{1}{3}}^\dagger}^2 c_{p+\frac{2}{3}}^\dagger c_{p+\frac{2}{3}}^2 c_{p-\frac{4}{3}} - \alpha' {c_p^\dagger}^3 c_{p+1} c_{p} c_{p-1} \} \tilde\Psi, \\
\end{eqnarray}
where $\tilde\Psi$ is a single permanent containing any number of violations of 102 or 111 clustering conditions, and $\alpha= t/V_{\text{\cancel{021}}}$,$\alpha' = t'/V_{\text{\cancel{3}}}$. 

In the model of the truncated Gaffnian Hamiltonian we presented above, the energy of the lowest lying neutral excitation significantly deviates from its classical prediction (which works accurately for Read-Rezayi states), but remains bounded by $\exp(-8\kappa^2/3)$. Nevertheless, one may wonder (1) if finite density of such excitations could imply gapless excitations in the thermodynamic limit, or if the inclusion of more terms in the Hamiltonian might lift the bound on the energy of the first excited states; and (2) whether the result may change if we consider a finite but fully translationally-invariant system (torus), instead of a cylinder. 

Regarding point (1), it appears likely that the Gaffnian remains gapped as we take the thermodynamic limit, as long as the Hamiltonian is truncated at the order assumed here. This is in agreement with a recent study based on perturbation theory~\cite{seidelunpub}. A more careful analysis is needed to understand the role of higher order terms and whether they affect the energy bound derived here. We suspect (2) is unlikely for the Hamiltonian defined in Eq.~(\ref{mr_trunc}). We note, however, that it might be of interest to consider an extension of the model to the order $\exp(-6\kappa^2)$, where three types of hopping terms arise: $021\leftrightarrow 102$, $1011\leftrightarrow 0120$ and $1002\leftrightarrow 0111$. Consider an example of 4 bosons on a torus; the root configuration for one of the excited states is $201100$. Using the mentioned hoppings, this configuration evolves into
\begin{eqnarray}
\nonumber 201100 \to 101011 \to 100120 \to 100201.
\end{eqnarray} 
However, the last configuration is exactly the translated version of the initial state. By forming linear combinations of such states, it might be possible to create exact zero-energy states in a translationally invariant system. However, due to the complexity of the Gaffnian Hamiltonian at truncation $\exp(-6\kappa^2)$, we have not found an explicit proof of this statement. 

\subsection{Haffnian}

As we go to more complicated states, the number of terms in the Hamiltonian quickly becomes intractable. For example, in the case of the Haffnian state of bosons at $\nu=1/2$~\cite{Read-PhysRevB.61.10267,green-10thesis, Hermanns-PhysRevB.83.241302}, related to an irrational CFT, we find the following terms relevant to the thin-torus expansion: 
\begin{eqnarray}\label{haff}
\nonumber && \text{\cancel{3}}, \text{\cancel{21}}, 030 \leftrightarrow 111, 102 \leftrightarrow 021, \text{\cancel{111}}, \text{\cancel{102}}, 0120 \leftrightarrow 1011, \\
\nonumber  && 0030 \leftrightarrow 1002, 0201 \leftrightarrow 1011, 0111 \leftrightarrow 1002, 00300 \leftrightarrow 10101, \\
\nonumber && \text{\cancel{1011}}, 00210 \leftrightarrow 10011, 111 \leftrightarrow 10101, 00120 \leftrightarrow 10002, \\
&&  01020 \leftrightarrow 10011, \text{\cancel{1002}}, 10020 \leftrightarrow 02001, \ldots 
\end{eqnarray}
Given the large number of terms in this case, we limit ourselves to classifying the Haffnian ground states in the thin-torus limit. It is known that Haffnian does not possess a well-defined ground-state degeneracy, but instead the manifold of degenerate ground states grows with the number of particles~\cite{Hermanns-PhysRevB.83.241302}. We would like to derive this microscopically, by studying the Hamiltonian Eq.~(\ref{haff}) near the thin torus limit.

The Haffnian has two simple thin-torus ground states that derive from root partitions $200200200...$ and $101010...$ (the latter one is shared with the Laughlin state). In addition to these classical patterns, we also find true quantum states where hoppings play a crucial role. We will explain the nature of these quantum states on the simplest example of 3 bosons on a torus. By taking into account translational symmetry, the Hilbert space consists of states $|1\rangle \equiv \widetilde{|100200\rangle}$, $|2\rangle \equiv \widetilde{|011100\rangle}$, and $|3\rangle \equiv \widetilde{|300000\rangle}$. It is implicitly understood that these states refer to the complete orbits of the translation operator, e.g., in zero-momentum sector $\widetilde{|10020\rangle}=1/\sqrt{3}(|100200\rangle + |001002\rangle + |020010\rangle)$, etc. The Haffnian Hamiltonian, Eq.~\ref{haff}, represented in this Hilbert space is given by 
\begin{equation}\label{haff_example}
\begin{bmatrix}
18H_{11} & 18 \sqrt{2} (2t') & 3 \sqrt{6} (2t'') \\
18 \sqrt{2} (2t') & 36 H_{22} & 6\sqrt{3}\tilde{t} \\
3\sqrt{6} (2t'') & 6\sqrt{3}\tilde{t} & 3H_{33}  
\end{bmatrix}
\end{equation}
Here the diagonal terms are given by $H_{11}=2V_{\text{\cancel{2001}}}+2t$, $H_{22}=V_{\text{\cancel{111}}}$ and $H_{33}=V_{\text{\cancel{3}}}$, and hopping terms are $t= V_{\text{\cancel{10020}} \leftrightarrow \text{\cancel{02001}}}$, $t'=V_{\text{\cancel{0111}}\leftrightarrow \text{\cancel{10002}}}$, $t''=V_{\text{\cancel{0030}}\leftrightarrow \text{\cancel{10002}}}$, and $\tilde{t}=V_{\text{\cancel{030}}\leftrightarrow \text{\cancel{111}}}$. Note the special role played by the hopping $t$ which hops the state 10020 $\leftrightarrow$ 02001 into itself on the torus, therefore giving a contribution to the \emph{diagonal} term $H_{11}$. Remarkably, the determinant of this matrix is zero, for \emph{any} value of $\kappa$. For reference, we quote the explicit expressions
of entries in the Eq.~(\ref{haff_example}):
\begin{eqnarray}\label{haff_example2}
\nonumber 18 H_{11} &=& 648 e^{-6 \kappa^2} (5-24 \kappa^2+72 \kappa^4), \\
\nonumber 18 \sqrt{2} (2t') &=& 648 \sqrt{2} e^{-4 \kappa^2} (5-16 \kappa^2+24 \kappa^4) \\
\nonumber 3 \sqrt{6} (2t'') &=& 18 \sqrt{6} e^{-3 \kappa^2} (5-12 \kappa^2) \\
\nonumber 36 H_{22} &=& 1296 e^{-2 \kappa^2} (5-8 \kappa^2+8 \kappa^4) \\
\nonumber 6\sqrt{3}\tilde{t} &=& 36 \sqrt{3} e^{-\kappa^2} (5-4 \kappa^2) \\
3H_{33}  &=& 15.
\end{eqnarray}

A non-trivial factorization property of the Haffnian Hamiltonian matrix elements for any $\kappa$ ensures that there are zero-energy ground states, for both even and odd number of particles, that cannot be expressed as single bosonic permanents. In order to count such configurations, we apply the following rule of a thumb: starting from the Laughlin root pattern, 1010101010, we can create an orthogonal state 0200101010, which violates \cancel{2001}. The energy of such a configuration can be brought back to zero using a combination of hoppings like in our example above, Eq.~\ref{haff_example}. Additional zero-energy states are obtained by creating more than one violation of \cancel{2001}. Combined with the regular, Laughlin-like root patterns, this gives a total degeneracy of $N+8$ or $N+1$, depending on the parity of the number of particles $N$~\cite{Hermanns-PhysRevB.83.241302}. We can also see that the same argument fails on the sphere or cylinder, because the hoppings eventually hit the ``boundary" and the energy remains above zero, resulting in a single ground state. Starting from the Haffnian ground state(s), one can construct excited states similarly to the Gaffnian case, but the analysis is more complicated due to a large number of terms in the Hamiltonian.

\section{States with spin}\label{sec:spin}

Finally, we also consider some non-unitary FQH states involving spin. In order to conform with the literature, in this section we discuss fermionic states, but similar analysis can be applied to bosons.

\subsection{Haldane-Rezayi}

The Haldane-Rezayi state of fermions at $\nu=1/2$ was initially proposed~\cite{Haldane-PhysRevLett.60.956} to describe the experimentally-observed quantized plateau at $\nu=5/2$, but was later identified as a critical (gapless) state of a $d$-wave superconductor with broken time reversal symmetry~\cite{Read-PhysRevB.61.10267,green-10thesis}. Ref.~\onlinecite{Seidel-PhysRevB.84.085122} discussed in detail the Haldane-Rezayi state in the limit of thin torus, arguing that it possesses gapless excitations. Here we provide a solvable model for the Haldane-Rezayi parent Hamiltonian and briefly analyze its solutions in the limit of thin torus and cylinder.

\begin{figure}[t]
\centerline{\includegraphics[width=\linewidth]{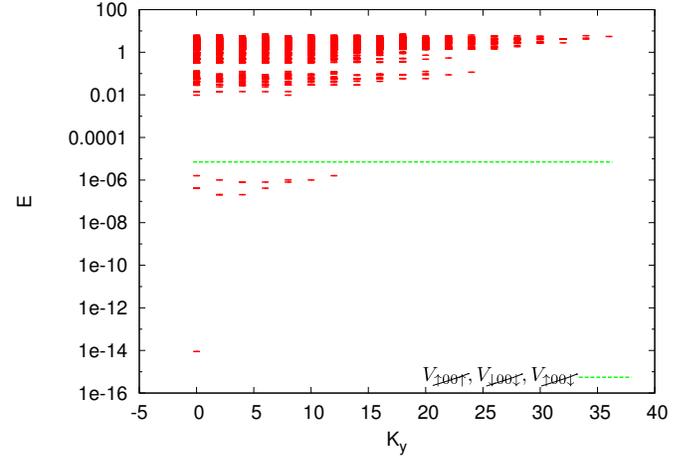}}
\caption[]{(Color online). Thin-cylinder energy spectrum (log scale) of the Haldane-Rezayi state for 6 fermions and 8 flux quanta, ratio 0.2. The spectrum includes all spin projections, but the low energy part belongs exclusively to the singlet sector. Line indicates the value of the density-density matrix elements $V_{\cancel{\uparrow 00 \uparrow}}$, $V_{\cancel{\downarrow 00 \downarrow}}$, and $V_{\cancel{\uparrow 00 \downarrow}}$. The excited states, however, have energies lower than this value, and ultimately become gapless when $L_y\to 0$.}
\label{fig_hr}
\end{figure}
The orbital part of the parent Hamiltonian of the Haldane-Rezayi state is the same as that of the Laughlin state -- just the $V_1$ Haldane pseudopotential~\cite{Haldane-PhysRevLett.60.956}. The presence of the spin degree of freedom allows for non-trivial dynamics to emerge in the thin-torus limit. The relevant terms in the Hamiltonian are the following:   
\begin{eqnarray}\label{hr_ham}\label{eq:hrterms}
\nonumber && \cancel{\uparrow\uparrow}, \cancel{\downarrow\downarrow}, \cancel{\uparrow\downarrow}; \; V_{\cancel{\uparrow\uparrow}} \sim e^{-\kappa^2/2}; \\
\nonumber && \cancel{\uparrow0\uparrow}, \cancel{\downarrow0\downarrow}, \cancel{\uparrow0\downarrow}; \; V_{\cancel{\uparrow0\uparrow}}  \sim 4 e^{-2\kappa^2}; \\
\nonumber  && \uparrow00\uparrow \leftrightarrow 0\uparrow\uparrow0, \downarrow00\downarrow \leftrightarrow 0\downarrow\downarrow0, \\
\nonumber && \uparrow00\downarrow \leftrightarrow 0\uparrow\downarrow0; \; t \sim 3 e^{-5\kappa^2/2}; \\
&& \cancel{\uparrow 00\uparrow}, \cancel{\downarrow 00\downarrow}, \cancel{\uparrow 00\downarrow}; \; V_{\cancel{\uparrow00\uparrow}} \sim 9e^{-9\kappa^2/2},
\end{eqnarray}
where the notation is the same as before, but each orbital can now be in a state 0,$\uparrow$, $\downarrow$, or $X=(\uparrow,\downarrow)$ (spin singlet). 

Let us first discuss the ground states of the Hamiltonian (\ref{hr_ham}) in the thin-torus limit. This Hamiltonian admits two types of simple ground states that are realized for even numbers of particles -- $X000X000...$ and $\underline{\uparrow\downarrow}00\underline{\uparrow\downarrow}00...$, where $\underline{\uparrow\downarrow}=\frac{1}{\sqrt{2}}(\uparrow\downarrow-\downarrow\uparrow)$ denotes a singlet formed in two adjacent orbitals. These ground states directly follow from the Hamiltonian (\ref{hr_ham}), however they do not exhaust all possible zero-energy solutions. More complicated ground states, that also exist for odd particle numbers, are allowed by the hopping term in Eq.~(\ref{hr_ham}). 

Let us again discuss the simplest case of 3 particles in order to illustrate how the energetics works out to ensure that zero-energy states exist. Assuming translation symmetry, the Hilbert space for 3 particles (zero-momentum sector) contains 5 states: 
\begin{eqnarray}
\nonumber && |\widetilde{\downarrow00X00\rangle},\widetilde{|0\uparrow\downarrow\downarrow00\rangle},\widetilde{|0\downarrow\downarrow\uparrow00\rangle},\\
&& \widetilde{|0\downarrow\uparrow\downarrow00\rangle},\widetilde{|\uparrow0\downarrow0\downarrow0\rangle}. 
\end{eqnarray}
For a general many-body state
\begin{eqnarray}
\nonumber \psi &=& |\widetilde{\downarrow00X00\rangle} + \alpha\widetilde{|0\uparrow\downarrow\downarrow00\rangle} + \beta\widetilde{|0\downarrow\downarrow\uparrow00\rangle}  \\
&& + \gamma \widetilde{|0\downarrow\uparrow\downarrow00\rangle} + \delta \widetilde{|\uparrow0\downarrow0\downarrow0\rangle},
\end{eqnarray}
we can compute the expectation value $\langle \psi|H\psi\rangle/\langle \psi|\psi\rangle$ w.r.t. to the Hamiltonian (\ref{hr_ham}) (note that this Hamiltonian is positive semidefinite):
\begin{eqnarray}
\nonumber E &\propto& 6 \gamma^2 (V_{\cancel{\uparrow\uparrow}} + V_{\cancel{\uparrow0\uparrow}}) +10 \delta^2 V_{\cancel{\uparrow0\uparrow}} + 3 \alpha^2 (3 V_{\cancel{\uparrow\uparrow}} + V_{\cancel{\uparrow0\uparrow}}) \\
\nonumber && +3 \beta^2 (3 V_{\cancel{\uparrow\uparrow}}+V_{\cancel{\uparrow0\uparrow}})+6 \beta (\gamma V_{\cancel{\uparrow\uparrow}}-3 t) \\
 && +6 \alpha (\gamma V_{\cancel{\uparrow\uparrow}}+3 t)+6\alpha \beta V_{\cancel{\uparrow0\uparrow}} + 9 V_{\cancel{\uparrow00\uparrow}}
\end{eqnarray}
It is easy to see that $\delta$ must be set to zero. Solving for $\alpha, \beta, \gamma$, we find that the zero-energy solution is obtained by choosing
\begin{eqnarray}
\alpha=-\beta=-\frac{t}{V_{\cancel{\uparrow\uparrow}}}, \gamma=0.
\end{eqnarray}
Therefore, for odd particle numbers, we always find a ground state that derives from a Slater determinant $|\downarrow00X00...\rangle$ containing an ``unpaired" electron, and configurations generated via pairwise hoppings. For even particle numbers, we can generate zero-energy states by the following process 
\begin{eqnarray}
0X0 \leftrightarrow (\uparrow0\downarrow - \downarrow0\uparrow).
\end{eqnarray}
Such a process comes with no energy penalty in the $V_1$ Hamiltonian. Using this process, starting from the root $X000X000X000...$, we can ``split" every doubly-occupied site $X$ and form a singlet in two next-nearest-neighbor orbitals. Such a singlet removes the energy penalty for $\uparrow0\downarrow$, while the electrostatic contribution from $...X00\uparrow...$ is cancelled by the hopping term, similarly to the above example. Such delocalized singlets were identified and discussed in Ref.~\cite{Seidel-PhysRevB.84.085122}.

On a finite cylinder, the mechanism above does not work and all the ground states are gapped out except for a single configuration, $X000X000...$. We are interested if some of the hopping processes studied above play an important role for the excited states. Inspired by the torus solution, one might try to construct the \emph{excited} state on the cylinder by hopping $0X0\leftrightarrow \uparrow0\downarrow$ somewhere in the interior of the system. However, it can be shown that such configurations, on a finite cylinder, must have an energy bounded by $V_{\cancel{\uparrow00\uparrow}}$. At the same order, only smaller due to a numerical prefact, we find the true first-excited state with an energy
\begin{equation}\label{hr_estimate}
6V_{\cancel{\uparrow00\uparrow}} - \frac{t^2}{V_{\cancel{\uparrow\uparrow}}} + \ldots,
\end{equation}
which belongs to the state $\uparrow\downarrow00X000X... - \downarrow\uparrow00X000X...$ (plus configurations obtained by hopping $\sigma00X \leftrightarrow 0\sigma\sigma'\bar{\sigma'}$, where $\sigma,\sigma'=\uparrow,\downarrow$ and $\bar{\sigma}$ is $\sigma$-flipped. Similarly to the Gaffnian case, we find that the classical estimate for the energies of the excited states of the Haldane-Rezayi state ($6V_{\cancel{\uparrow00\uparrow}}$) is an overestimate due to the hopping contribution. The quantum correction in Eq.~(\ref{hr_estimate}) agrees very accurately with exact diagonalization results, Fig.~\ref{fig_hr}. 
\begin{figure}[t]
\centerline{\includegraphics[width=\linewidth]{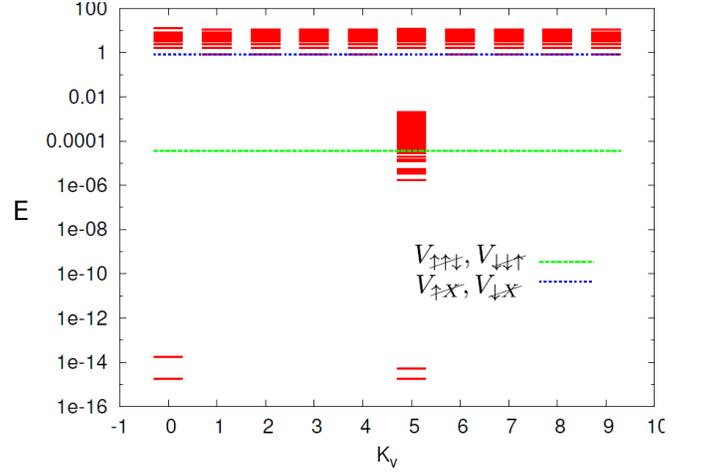}}
\caption[]{(Color online). Energy spectrum (log scale) of the full permanent Hamuiltonian at $\nu=1$ for 10 fermions and 10 flux quanta, torus aspect ratio 1/12. The spectrum includes all spin projections. Lines indicate the values of the corresponding density-density matrix elements of the Hamiltonian, which are expected to define the energy bands of the truncated Hamiltonian (\ref{perm_ham}). For the purpose of clarity, spectrum is only plotted versus $K_y$ quantum number. Because $N_e$ is even, sector $(5,5)$ corresponds to the zero momentum (center of the Brillouin zone), where the ground state has extensive degeneracy.}
\label{fig_perm}
\end{figure}

\subsection{Permanent state}

Yet another critical state, introduced in Ref.~\onlinecite{Read-PhysRevB.54.16864}, is the permanent state. In the simplest case describing $\nu=1$ fermions, it is given by a 3-body Hamiltonian penalizing the closest possible approach of three spin-1/2 fermions. It was argued that this state was critical and at a phase transition from ferromagnet to paramagnet~\cite{Read-PhysRevB.54.16864}. 

Permanent Hamiltonian on the torus has a single, spin-singlet zero-energy ground state in sectors corresponding to the corner of the Brillouin zone and the midpoints of the two sides, i.e. for Haldane pseudomomenta $\mathbf{k}=(N/2,0)$, $(0,N/2)$ and $(N/2,N/2)$. On the other hand, in a $\mathbf{k}=0$ sector, the zero-energy ground-state in fact exists for any projection of total spin $S=N/2$, leading to a macroscopic degeneracy. In the thin torus limit, as we see below,  there are also additional zero modes that ``sink" through the $\mathbf{k}=0$ sector. 

The permanent Hamiltonian has no penalty for scattering $\uparrow\uparrow\uparrow$ into $\uparrow\uparrow\uparrow$ (nor $\downarrow\downarrow\downarrow$ into $\downarrow\downarrow\downarrow$), and the relevant terms only involve two $\uparrow$ and one $\downarrow$ particle at a time (or two $\downarrow$ and one $\uparrow$). In the thin torus limit, the orbital part of the permanent Hamiltonian is expanded as follows
\begin{eqnarray}\label{perm_ham}
\nonumber && \cancel{\uparrow X}, \cancel{\downarrow X} \sim e^{-2\kappa^2/3},\\
\nonumber && 0X\uparrow \leftrightarrow \uparrow 0X, 0X\downarrow \leftrightarrow \downarrow 0X \sim e^{-5\kappa^2/3}, \\
\nonumber && \cancel{\uparrow \downarrow \uparrow}, \cancel{\downarrow \uparrow \downarrow} \sim 4e^{-2\kappa^2},\label{eq:permanentterms}\\
\nonumber && \downarrow\uparrow\uparrow \leftrightarrow \uparrow\downarrow\uparrow, \uparrow\downarrow\downarrow \leftrightarrow \downarrow\uparrow\downarrow \sim 2e^{-2\kappa^2},\\
\nonumber  &&  \cancel{\downarrow\uparrow\uparrow}, \cancel{\uparrow\downarrow\downarrow} \sim e^{-2\kappa^2},\\
 && \ldots 
\end{eqnarray} 
The excitation spectrum in the thin torus (Fig.~\ref{fig_perm}) has a peculiar structure -- its entire low-energy part is built out of states whose momentum has at least one component equal to zero. In the spin-singlet sector, unique zero-energy ground state is found for any aspect ratio in each of the four high-symmetry points in the Brillouin zone. Two of those ground states are associated with the pattern $X0X0...$, which is also the unique ground
state in systems with a ``boundary" (cylinder or sphere). In these sectors, the spectrum has a gap equal to the $\cancel{\uparrow X}$ density-density term. 

In sectors with $k_y=0$, we find the additional two zero-energy ground-states that are not a single Slater determinant. They contain the spin-separated state $\uparrow\uparrow\ldots\uparrow\downarrow\ldots \downarrow\downarrow$, and the configurations obtained from it by hopping around the domain wall $\downarrow\uparrow\uparrow \leftrightarrow \uparrow\downarrow\uparrow$. With the exception of two ground states $X0X0...$, the low-energy manifold resides in a restricted Hilbert space where double occupancy of a single orbital is forbidden. At $\nu=1$, this implies that no orbital is empty, either. The effective Hamiltonian in this restricted space contains only the last three types of terms in Eq.~\ref{perm_ham}, and the system maps onto the Halperin-111 state. The ground-state of the permanent in the sector $\mathbf{k}=0$ reduces to the Halperin-111 state for large $\kappa$, and the excited states overlap with 99.7\%. Therefore, in the thin torus limit, the low-lying spectrum of the permanent includes the 111 state and its gapless excitations, which are well-known in the literature~\cite{Seidel-PhysRevLett.101.036804}. 

\section{Conclusions}\label{sec:conc}

In this work we have developed a method to study the properties of many-particle repulsive Hamiltonians that define various quantum Hall model states on the torus and cylinder geometries. The obtained form of these Hamiltonians allows for transparent perturbative expansions in terms of $\kappa$, that enables analytical treatments and casts these models in light of Hubbard-type models known in other areas of physics, but with generalized types of hopping processes that involve clusters of $k$ particles while preserving their center of mass momentum. 

Using the above tools, we have addressed several specific physical questions. For example, how do the members of the Read-Rezayi sequence differ from one another, and do they all have similar solvable limits, previously known for the Laughlin case. We have shown that they indeed remain gapped and behave classically as $L_y\to 0$, as one would expect for the states based on unitary CFTs. The Read-Rezayi ground states, as well as the entire neutral excitation spectrum, can be classified by the clustering properties enforced by the parent Hamiltonians as $L_y\to 0$. Corrections that build in quantum fluctuations can be analytically obtained in some cases for small but finite $L_y$. We must emphasize that the statements about gaps of various states have been derived near the 1D limit, and do not directly apply to the isotropic (2D) limit. 

Another physical question that naturally arises in this context is whether the thin-torus expansion can reveal in some transparent way the difference between states whose underlying CFT is rational and unitary, as opposed to the ones described by non-unitary or irrational CFT. Although we do not have a general proof of this fact, our analysis has identified a general presence of hopping terms in the Hamiltonians of non-unitary and irrational states that lead to inherently quantum-mechanical behavior in the thin-torus limit. Further study of individual cases, using the models derived here as well as from a large family of Hamiltonians proposed recently in Ref.~\onlinecite{jackson}, would be needed to rigorously prove the existence of gapless excitations in thermodynamic limit~\cite{gaffnian_new1, gaffnian_new2} (or lack thereof). Again, such conclusions would not automatically hold in the isotropic (2D) limit, but they could potentially serve as a useful diagnostic to apply to states whose nature is a priori not known.   

One distinct advantage of the method presented here is that is generic and can be applied to any given Hamiltonian. As such, it might be useful for generating approximate matrix-product state (MPS) expressions for any type of quantum Hall state, given that alternative methods~\cite{Zaletel-PhysRevB.86.245305,Estienne-PhysRevB.87.161112} depend sensitively on the type of CFT. Unfortunately, it is known that MPS based on the first-order thin torus expansion~\cite{Jansen-2012JMP....53l3306J,Soule-PhysRevB.85.155116,Nakamura-PhysRevLett.109.016401,Wang-PhysRevB.87.245119} is quite different from the alternate one based on CFT, and yields a poor description of the state in the isotropic (2D) limit~\cite{Estienne-PhysRevB.87.161112}. An open question remains whether higher-order corrections near the thin-torus limit can generally be organized in a tractable way to achieve an accurate MPS for $L_y \gg \ell_B$, perhaps in a form of generalized Schrieffer-Wolff~\cite{sw} or continuous unitary transformations~\cite{cut} that have been applied  successfully to the Hubbard model.       

\section*{Acknowledgements}

I am indebted to B. Andrei Bernevig for originally suggesting this problem. I would like to thank him and N. Regnault for numerous helpful comments and critical reading of an earlier version of the manuscript. I also acknowledge fruitful discussions with F. Essler and thank N. Read for clarifying some properties of the Haffnian CFT. This work was supported by DOE grant DE-SC$0002140$.  Research at Perimeter Institute is supported by the Government of Canada through Industry Canada and by the Province of Ontario through the Ministry of Economic Development \& Innovation.

\appendix
\section{Derivation of matrix elements}\label{sec_app}

Here we demonstrate the derivation of a convenient form of the Hamiltonian matrix elements for the Laughlin state on the torus using the Poisson summation formula. 

A general, translationally-invariant two-body Hamiltonian in the second-quantized notation is given by
\begin{equation}\label{2bodyrhoq}
H=\sum_{\mathbf{q}} V(\mathbf{q}) \rho_\mathbf{q} \rho_{-\mathbf{q}},
\end{equation}
where in our case the sum runs over discrete momenta $\mathbf{q}=(q_x, q_y) = (2 \pi s/L_x, 2 \pi t/L_y)$ in the Brillouin zone (assumed to be rectangular, for simplicity),
$V(\mathbf{q})$ is the Fourier transform of the (two-body) interaction, 
and $\rho_\mathbf{q}$ is the Fourier component of the density operator,
\begin{equation}\label{rhoq}
\rho_\mathbf{q} = \sum_{j,j'} \langle j|e^{i\mathbf{q}\mathbf{r}}|j' \rangle c_j^\dagger c_{j'}.
\end{equation}
Here $c_j^\dagger$ creates a particle in the state $|j\rangle$ given in Eq.~(\ref{eq:onebodywf}). After some algebra, $\rho_\mathbf{q}$ can be written as
\begin{equation}\label{rhoqev}
\rho_\mathbf{q} = e^{-|\mathbf{q}|^2/4} \sum_{j=0}^{N_\phi-1} e^{-i\frac{2\pi}{N_\phi} s(j+t/2)} c_j^\dagger c_{j+t}.
\end{equation}
Using this result, we can rewrite the Hamiltonian (\ref{2bodyrhoq}) as
\begin{equation}\label{2bodygen}
H=\sum_{j_1,...,j_4} V_{j_1j_2j_3j_4} c_{j_1}^\dagger c_{j_2}^\dagger c_{j_3} c_{j_4},
\end{equation}
where $V_{j_1j_2j_3j_4}$ are given by~\cite{chak}
\begin{eqnarray}
\nonumber V_{j_1j_2j_3j_4} = \sum_{(s,t) \neq (0,0)} && \frac{1}{2L_xL_y} V(\mathbf{q})  \label{standard}\\
 \nonumber && \times e^{-q_x^2/2 -q_y^2/2 - i q_x (X_{j_3}-X_{j_4})} \\
\nonumber && \times \sum_{\tilde{t}} \delta_{q_y, X_{j_1}-X_{j_4}+\tilde{t}L_x} \\
&& \sum_l \delta_{X_{j_1}+X_{j_2},X_{j_3}+X_{j_4}+lL_x}
\end{eqnarray} 
where $X_j =  2\pi j/ L_y$. Different interactions can be studied by simply redefining $V(\mathbf{q})$; for example, if we are interested in the Laughlin state of bosons, we should define $V(\mathbf{q})$ to be any positive constant, e.g. $V(\mathbf{q})=1$. The disadvantage of Eq.~(\ref{standard}) is that we still need to evaluate a double nested sum over $s$ and $t$.

We can however rewrite the sum over $s$ (i.e., $q_x$) using the Poisson summation formula
\begin{equation}
\sum_{n=-\infty}^{\infty} f(n) = \sum_{\tilde{n}=-\infty}^{\infty} \int_{-\infty}^{\infty} f(x) e^{-i2\pi \tilde{n} x} dx,\label{eq:Posson}
\end{equation}
to obtain, in the case of the bosonic Laughlin state with $V(q)=1$:
\begin{eqnarray}
\nonumber V_{j_1j_2j_3j_4} \propto \sum_l \delta_{X_{j_1}+X_{j_2},X_{j_3}+X_{j_4}+lL_x} \\
\sum_{\tilde{s},\tilde{t}} e^{-\frac{1}{2}\left( X_{j_1} - X_{j_3} + \tilde{s}L_x \right)^2 -\frac{1}{2}\left( X_{j_1} - X_{j_4} + \tilde{t}L_x \right)^2}.
\end{eqnarray}
Using the identity 
\begin{eqnarray}
\nonumber && 2(a_1+b_1)^2+2(a_2+b_2)^2 = \\
&& \left[ (a_1+a_2) + (b_1+b_2) \right]^2
 + \left[ (a_1-a_2) + (b_1-b_2) \right]^2, \label{identity}
\end{eqnarray}
we can switch from sums over integer $\tilde{s},\tilde{t}$ to sums over $c\equiv \tilde{s} + \tilde{t}, r\equiv \tilde{s} - \tilde{t}$. For consistency, we must only sum over $c$ and $r$ of the same parity. We finally obtain
\begin{eqnarray}\label{bosoniclaughlin}
&& V_{j_1j_2j_3j_4} \propto \sum_l \delta_{X_{j_1}+X_{j_2},X_{j_3}+X_{j_4}+lL_x} \\
&&\nonumber \Big\{ \sum_{c \in {\rm even}} e^{-\frac{1}{4} \left( X_{j_1} - X_{j_2} + lL_x + cL_x \right)^2 } \sum_{r \in {\rm even}} e^{-\frac{1}{4} \left( X_{j_3} - X_{j_4} + rL_x \right)^2 }  \\
&&\nonumber+ \sum_{c \in {\rm odd}} e^{-\frac{1}{4} \left( X_{j_1} - X_{j_2} + lL_x + cL_x \right)^2 } \sum_{r \in {\rm odd}} e^{-\frac{1}{4} \left( X_{j_3} - X_{j_4} + L_xa \right)^2 } \Big\} 
\end{eqnarray}
Therefore, the Poisson summation formula has simplified each matrix element by breaking it into a product of two sums. Each of the sums is a function only of the indices 
of the creation \emph{or} annihilation operators, contrary to Eq.~(\ref{standard}) that mixes the two groups of indices. Note that this form of the matrix element also becomes similar to other geometries, like the disk or the sphere, where the Hamiltonian can be decomposed in sums over two-boson creation and annihilation clusters~\cite{Chandran-PhysRevB.84.205136}. The periodic boundary condition is reflected through the constraint that sums over $c$ and $r$ must be performed over only even or only odd integers. 

We mention that the summation formula will be effective for any short-range two-body interaction, but not for Coulomb because the corresponding integral cannot be evaluated in closed form. In practice, computing the matrix elements for two-body interactions can be achieved with little cost even without resorting to the Poisson formula, however for higher order interactions (3-body, etc.), there will be an important speedup. For example, in the case of 3-body interactions that are needed for the bosonic Gaffnian state~\cite{Simon-PhysRevB.75.075318},
\begin{equation}
H_{\rm Gaff} = \sum_{i<j<k}\mathcal{S}_{ijk} \{ \nabla_i^4 \delta(\mathbf{r}_i-\mathbf{r}_j) \delta(\mathbf{r}_j-\mathbf{r}_k) \}.
\end{equation}
where $\mathcal{S}$ is a symmetrizer, by brute force one would need to compute 4 nested sums. Instead, we can derive the following equivalent expression for the Gaffnian matrix elements which is significantly faster:
\begin{eqnarray}
\nonumber V_{j_1\ldots j_6} = 3 \sum_{g=0,1,2} \sum_{k_r,k_s = g \; {\rm mod} \; 3}  e^{-A} \sum_{k_r',k_s' = l+g \; {\rm mod} \; 3} 
e^{-A'} \\
\nonumber - 2 \sum_{g=0,1,2} \sum_{k_r,k_s = g \; {\rm mod} \; 3} A e^{-A} \sum_{k_r',k_s' = l+g \; {\rm mod} \; 3} e^{-A'} \\
\nonumber - 2 \sum_{g=0,1,2} \sum_{k_r,k_s = g \; {\rm mod} \; 3} e^{-A} \sum_{k_r',k_s' = l+g \; {\rm mod} \; 3} A' e^{-A'} \\
\nonumber + 4 \sum_{g=0,1,2} \sum_{k_r,k_s = g \; {\rm mod} \; 3} A e^{-A} \sum_{k_r',k_s' = l+g \; {\rm mod} \; 3} A' e^{-A'}
\end{eqnarray}
where $A= \tilde{X}_r^2+\tilde{X}_r\tilde{X}_s+\tilde{X}_s^2 $ and $A' = \tilde{X}_{r'}^2+\tilde{X}_{r'}\tilde{X}_{s'}+\tilde{X}_{s'}^2$, $l$ denotes the momentum transfer, $(j_6+j_5+j_4-j_3-j_2-j_1)/N_\phi$, and 
\begin{eqnarray}
\nonumber \tilde{X}_r = (2 X_{j_1} - X_{j_2} - X_{j_3} + k_r L_x)/3 \\
\nonumber \tilde{X}_s = (2 X_{j_2} - X_{j_1} - X_{j_3} + k_s L_x)/3
\end{eqnarray}
(and similarly for $\tilde{X}_{r'},\tilde{X}_{s'}$ using $k_r',k_s'$).

With an overall factor $(2\sqrt{\pi}/3)\sqrt{3\pi}/L_y^2$, the above Hamiltonian is correctly normalized to yield energies 0, 1 and 2 for 3 particles in the thermodynamic limit.

\section{Factorization property of Read-Rezayi Hamiltonians}\label{sec:factorization}

In the Laughlin case (Sec.~\ref{sec:laughlin}), we have seen that by truncating the Hamiltonian at the order $\exp(-2\kappa^2)$, it was possible to reexpress it as a positive semidefinite operator. Here we analyze in more details if such a factorization is still possible when we continue the expansion further in $\kappa$.

In order to express the truncated Laughlin Hamiltonian as $\sum_p A_p^\dagger A_p$, we seek $A_p^\dagger$ as a sum of terms $c_{p+r}^\dagger c_{p-r}^\dagger$. Each such term contributes $r^2$ to the exponent in the total matrix element, therefore the possible terms in $A_p^\dagger$ can be easily classified according to the increasing value of $r$:
\begin{eqnarray}\label{laughlin_fact}
\nonumber && {c_p^\dagger}^2, \; r^2=0, \\
\nonumber && c_{p+\frac{1}{2}}^\dagger {c_{p-\frac{1}{2}}^\dagger}, \; r^2=1/4, \\
\nonumber && c_{p+1}^\dagger {c_{p-1}^\dagger}, \; r^2=1, \\
\nonumber && c_{p+\frac{3}{2}}^\dagger {c_{p-\frac{3}{2}}^\dagger}, \; r^2=9/4, \\
\nonumber && c_{p+2}^\dagger {c_{p-2}^\dagger}, \; r^2=4, \\
&& \ldots
\end{eqnarray}
By grouping these terms we recover the full Hamiltonian, whose dominant terms were given in Eq.~(\ref{laughlin_trunc}). Clearly, this can be done because each matrix element is a simple Gaussian which can be factorized $e^{A_1+A_2}=e^{A_1}e^{A_2}$. Furthermore, we see that a positive semidefinite form can be obtained if we stop at the order $r=1$. At this order, the smallest term will be the density-density term $(c_{p+1}^\dagger {c_{p-1}^\dagger})(c_{p-1}c_{p+1})$, which has a combined weight $\exp(-2\kappa^2)$. Keeping the first three terms in Eq.~(\ref{laughlin_fact}) is guaranteed to reproduce all the terms of the full Hamiltonian down to $\exp(-2\kappa^2)$. This is easily seen because the inclusion of any of the terms with $r>1$ in Eq.~(\ref{laughlin_fact}) would give rise to terms in the Hamiltonian with weights smaller than $\exp(-9\kappa^2/4)$, which is below $\exp(-2\kappa^2)$. Therefore, by keeping the first three terms in Eq.~(\ref{laughlin_fact}) our expression completely closes. 

\begin{figure}[htb]
\centerline{\includegraphics[width=\linewidth]{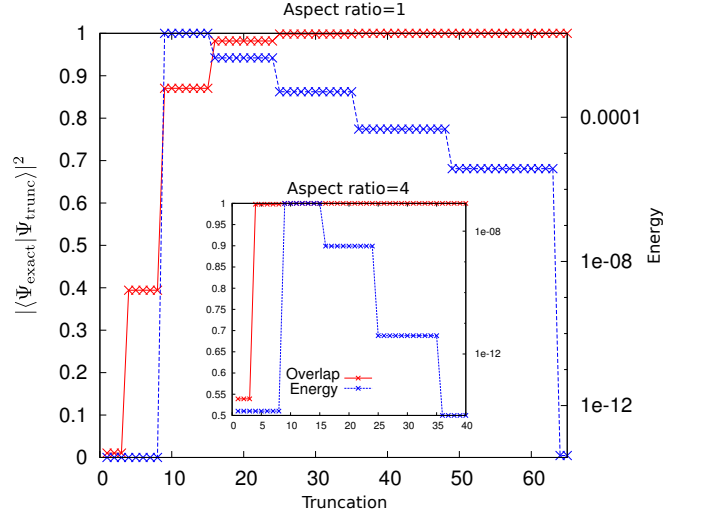}}
\caption[]{(Color online). Comparison between the full Laughlin state and the ground state of the truncated Hamiltonian $H=\sum_p A_p^\dagger A_p$, as a function of truncation in $A_p^\dagger$ (i.e. $4r^2$ in Eq.(~\ref{laughlin_fact})). Left axis shows the overlap between the ground states of the truncated and the full Hamiltonian, and the right axis shows the ground-state energy of the truncated Hamiltonian. Data is for $N=8$ particles on an isotropic torus (aspect ratio 1), and near the thin torus (aspect ratio 4, inset).}
\label{fig_laughlin_truncspectrum}
\end{figure}

Additionally, we see that a closed expression cannot be obtained by keeping more terms in $\kappa$. Let us try to go to the next order by keeping terms down to $\exp(-9\kappa^2/4)$ in Eq.~(\ref{laughlin_fact}). This will constitute an approximation to the full Hamiltonian at the order $\exp(-9\kappa^2/2)$, with the smallest term $(c_{p+3/2}^\dagger {c_{p-3/2}^\dagger})(c_{p-3/2}c_{p+3/2})$. In doing so, we have missed a hopping term  that combines $c_{p+2}^\dagger c_{p-2}^\dagger$ with $c_p^2$. The weight of this term is equal to $\exp(-4\kappa^2)$, thus it is \emph{larger} than $\exp(-9\kappa^2/2)$ and should be present in the expansion. This implies that we must augment our definition of $A_p^\dagger$ with also the next term, $c_{p+2}^\dagger c_{p-2}^\dagger$. Unfortunately, this does not resolve the issue. 

Generally, if we construct an approximation to $A_p^\dagger$ by keeping terms with the exponent $\leq r_0^2$, we are hoping to recover all the terms in the Hamiltonian $\geq \exp(-2r_0^2\kappa^2)$. However, we will necessarily miss a term with the weight $\exp(-(r_0+1/2)^2\kappa^2)$, for $r_0$ half-integer, or term $\exp(-(r_0+1/2)^2\kappa^2-1/4\kappa^2)$, for $r_0$-integer. By comparing the competing terms, we get a simple condition for expressing the Hamiltonian in form $A_p^\dagger A_p$:
\begin{equation}
(r_0-\frac{1}{2})^2 < \frac{3}{4} \;\;\; {\rm or} \;\;\; \frac{1}{2},
\end{equation}
for $p$ integer or half-integer, respectively. It is easy to see that this is satisfied only for small $r_0$ like 1/2 and 1.  The problem illustrated on this particular example only becomes worse with the increase of the number of terms in the expansion, and as we go to more complicated states (Moore-Read and Read-Rezayi).

In conclusion, if we truncate the Hamiltonian at some order in $\kappa$, in general we will not be able to exactly rewrite it as $\sum_p A_p^\dagger A_p$, except at a very low order of the expansion. However, we can proceed in a slightly different fashion and work directly with $A_p^\dagger$ and its expansion, like we did in Eq.~\ref{laughlin_fact}. By expanding $A_p^\dagger$ to the order $\exp(-\alpha \kappa^2)$, we generate a positive semidefinite operator $\tilde H = \sum_p A_p^\dagger A_p$ that ``approximates" the full Hamiltonian $H$ to the order $\exp(-2\alpha\kappa^2)$, in the sense that the eigenstates of $\tilde H$ have large overlap with those of $H$. This is verified numerically in Fig.~\ref{fig_laughlin_truncspectrum} where we plot the overlap of the ground state of $\tilde H$ with the full Laughlin state, as a function of the order of truncation in $A_p^\dagger$. The ``truncation" $\Sigma$ means that we keep only terms $c_{p+r}^\dagger c_{p-r}^\dagger$ in the expansion of $A_p^\dagger$ such that $4r^2\leq \Sigma$. Thus, as we vary $\Sigma$ over all integers, we expect to see plateaus in the overlap or ground-state energy because new interaction terms in the Hamiltonian appear in discrete steps $4r^2=0, 1, 4, 9, 16, \ldots$.  

In Fig.~\ref{fig_laughlin_truncspectrum} (inset), the overlap monotonically approaches unity, and does so much faster for aspect ratio closer to the thin-torus limit. Unlike the overlap, the ground-state energy has a non-monotonic dependence on the order of truncation. For $4r^2\leq 4$, it is zero to machine precision (as we expect from the analytic solution), but for $4r^2=9$ it jumps to a non-zero value. We have analytically verified that this occurs for $N\geq 4$ particles (for $N=3$ one can show analytically that the solution for $4r^2\leq 9$ also has zero energy, i.e. it yields the full Laughlin state for that finite system). After this jump, the energy monotonically decays to zero, but interestingly this ``relaxation" occurs much more slowly compared to the saturation of the overlap. 

\begin{figure}[htb]
\centerline{\includegraphics[width=\linewidth]{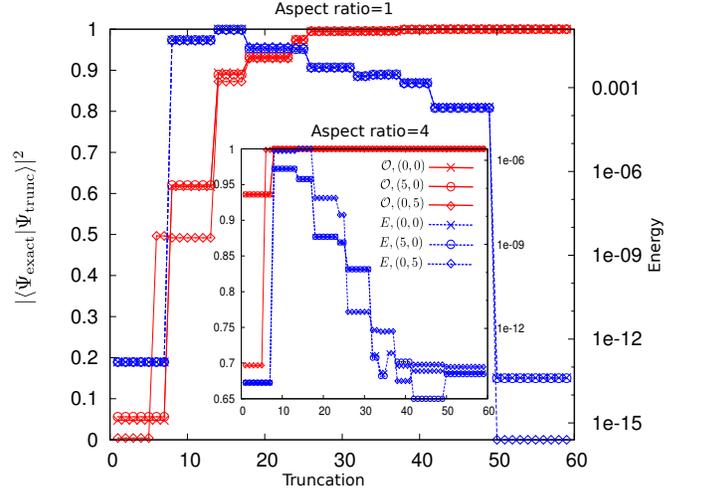}}
\caption[]{(Color online). Comparison between the full Moore-Read state and the ground state of the truncated Hamiltonian $\tilde H$, as a function of truncation i.e. $6\Sigma$ in Eq.(~\ref{mr_fact})). Left axis shows the overlap between the ground states of the truncated and the full Hamiltonian, and the right axis shows the ground-state energy of the truncated Hamiltonian, for each of the topologically-degenerate sectors labelled by (0,5), (5,0), (0,0). Data is for $N=10$ particles on an isotropic torus (aspect ratio 1), and near the thin torus (aspect ratio 4, inset).}
\label{fig_mr_truncspectrum}
\end{figure}   

In the bosonic Moore-Read case, the operator $A_p^\dagger$ (previously given in Eq.~(\ref{laughlin_fact}) for the Laughlin case) generalizes to a sum of terms of the form $c_{p+r}^\dagger c_{p+s}^\dagger c_{p-r-s}^\dagger$. Each such term contributes $\Sigma=r^2+s^2+rs$ to the total weight of the matrix element in the Hamiltonian. As we explained in Sec.~\ref{sec:approach}, $r,s$ here take values in $\mathbb{Z}/3$, which yields the following possibilities for $c_{p+r}^\dagger c_{p+s}^\dagger c_{p-r-s}^\dagger$ in the order of increasing $\Sigma$:
\begin{eqnarray}\label{mr_fact}
\nonumber && r=0, \; s=0, \; -r-s=0; \; \Sigma=0 \\
\nonumber && r=1/3, \; s=1/3, \; -r-s=-2/3; \; \Sigma=1/3 \\
\nonumber && r=1, \; s=0, \; -r-s=-1; \; \Sigma=1 \\
\nonumber && r=2/3, \; s=2/3, \; -r-s=-4/3; \; \Sigma=4/3 \\
\nonumber && r=4/3, \; s=1/3, \; -r-s=-5/3; \; \Sigma=7/3 \\
\nonumber && r=2, \; s=-1, \; -r-s=-1; \; \Sigma=3 \\
\nonumber && r=2, \; s=0, \; -r-s=-2; \; \Sigma=4 \\
&& \ldots
\end{eqnarray}
It is clear that by keeping only the first two terms we do not generate any hoppings but only density-density terms (because ${c_p^\dagger}^3$ and $c_{p+1/3}^2c_{p-2/3}$ cannot be combined together, since $p$ must be either an integer \emph{or} a fraction). Thus, with these two terms we recover the strict thin-torus limit. 

More interestingly, to obtain the correction to the thin-torus limit, we keep the terms in Eq.~(\ref{mr_fact}) up to $\Sigma=4/3$. One might expect this would reproduce the full Moore-Read Hamiltonian to the order $\exp(-8\kappa^2/3)$. This is almost true, apart from a single type of term
\begin{equation}\label{neglect}
c_{p+4/3}^\dagger c_{p+1/3}^\dagger c_{p-5/3}^\dagger c_{p+1/3}^2 c_{p-2/3}.
\end{equation}
The order of this term is $\exp(-8\kappa^2/3)$ but it cannot be reproduced by keeping terms in $A_p^\dagger$ to the order $\exp(-4\kappa^2/3)$. In order to construct a solvable model, we must neglect this term. In the vicinity of the thin-torus limit, however, neglecting this term produces a negligible error. 

In Fig.~\ref{fig_mr_truncspectrum}, similarly to the Laughlin case, we assess the quality of the approximation achieved by truncating the Hamiltonian via Eq.(~\ref{mr_fact}). We compare the ground state of the full Moore-Read Hamiltonian and the truncated Hamiltonian obtained by keeping only terms $c_{p+r}^\dagger c_{p+s}^\dagger c_{p-r-s}^\dagger$ with $\Sigma$ smaller than a given cutoff. We evaluate the overlap between the two ground states in each of the topologically-degenerate sectors, as well as the ground-state energy of the truncated Hamiltonian. The overlap monotonically approaches unity with the increase of the cutoff, while the ground-state energy first increases and then slowly relaxes down to zero. We see that, in general, the truncation of the Hamiltonian introduces a splitting between the ground-state energy in different topological sectors, although the ground states monotonically evolve towards the full Moore-Read states as the truncated order is increased. The location of the pleateaus in Fig.~\ref{fig_mr_truncspectrum} can be traced back to the special values of $\Sigma$ in Eq.(~\ref{mr_fact}).

Finally, in the $\mathbb{Z}_3$ RR case, $A_p^\dagger$ is constructed from terms of the form $c_{p+r}^\dagger c_{p+s}^\dagger c_{p+t}^\dagger c_{p-r-s-t}^\dagger$. Here $r,s,t$ take values $\mathbb{Z}/4$, and each term contributes $\Sigma=r^2+s^2+t^2+rs+rt+st$ to the total weight, therefore they can be classified in the order of dominance:
\begin{eqnarray}\label{rr_fact}
\nonumber && r=0, \; s=0, \; t=0; \; -r-s-t=0; \; \Sigma=0 \\
\nonumber && r=1/4, \; s=1/4, \; t=1/4; \; -r-s-t=-3/4; \; \Sigma=3/8 \\
\nonumber && r=1/2, \; s=1/2, \; t=-1/2; \; -r-s-t=-1/2; \; \Sigma=1/2 \\
\nonumber && r=1, \; s=0, \; t=0; \; -r-s-t=-1; \; \Sigma=1 \\
&& \ldots
\end{eqnarray}
In order to reproduce the thin-torus ground states, we must keep terms up to $\Sigma=1/2$ in Eq.~(\ref{rr_fact}). This will generate an approximation to the full RR Hamiltonian at the order of $\exp(-\kappa^2)$, because it misses the term which combines $\Sigma=1$ with $\Sigma=0$, e.g. 
\begin{equation}\label{rrhop}
c_{p+1}^\dagger {c_p^\dagger}^2 c_{p-1}^\dagger c_p^4.
\end{equation}
This is different from the previous cases (Laughlin and Moore-Read) because the ``minimal" Hamiltonian that describes the thin-torus limit is now no longer ``protected" from hopping terms, i.e., it becomes intrinsically \emph{non-classical} and its properties, such as the existence of a gap, become less obvious.

Finally, we mention in passing that further approximations in $\kappa$ can be generated for the full RR Hamiltonian by keeping more terms in $A_p^\dagger$, Eq.~(\ref{rr_fact}), similarly to the Moore-Read state (Sec.~\ref{sec:mr}). For example, by keeping terms down to $\Sigma=1$ in Eq.~(\ref{rr_fact}) we obtain an exactly-solvable Hamiltonian that approximates the full RR Hamiltonian at the order $\exp(-2\kappa^2)$. Strictly speaking, this solvable Hamiltonian is an approximation to the full Hamiltonian because it misses some terms at a given order: e.g., ${c_{p+1}^\dagger}^2 {c_{p-1}^\dagger}^2 c_p^4$ has a weight $\exp(-2\kappa^2)$, therefore it is present in the full Hamiltonian but cannot be obtained from $A_p^\dagger$ truncated to order $\Sigma=1$, etc.

\bibliography{thintorus}

%Merlin.mbs v4.21 2009-07-09.
\begin{thebibliography}{10}%
\makeatletter
\providecommand \@ifxundefined [1]{%
 \ifx #1\undefined \expandafter \@firstoftwo
 \else \expandafter \@secondoftwo
\fi
}%
\providecommand \@ifnum [1]{%
 \ifnum #1\expandafter \@firstoftwo
 \else \expandafter \@secondoftwo
\fi
}%
\providecommand \enquote [1]{``#1''}%
\providecommand \bibnamefont  [1]{#1}%
\providecommand \bibfnamefont [1]{#1}%
\providecommand \citenamefont [1]{#1}%
\providecommand\href[0]{\@sanitize\@href}%
\providecommand\@href[1]{\endgroup\@@startlink{#1}\endgroup\@@href}%
\providecommand\@@href[1]{#1\@@endlink}%
\providecommand \@sanitize [0]{\begingroup\catcode`\&12\catcode`\#12\relax}%
\@ifxundefined \pdfoutput {\@firstoftwo}{%
 \@ifnum{\z@=\pdfoutput}{\@firstoftwo}{\@secondoftwo}%
}{%
 \providecommand\@@startlink[1]{\leavevmode\special{html:<a href="#1">}}%
 \providecommand\@@endlink[0]{\special{html:</a>}}%
}{%
 \providecommand\@@startlink[1]{%
  \leavevmode
  \pdfstartlink
   attr{/Border[0 0 1 ]/H/I/C[0 1 1]}%
   user{/Subtype/Link/A<</Type/Action/S/URI/URI(#1)>>}%
  \relax
 }%
 \providecommand\@@endlink[0]{\pdfendlink}%
}%
\providecommand \url  [0]{\begingroup\@sanitize \@url }%
\providecommand \@url [1]{\endgroup\@href {#1}{\urlprefix}}%
\providecommand \urlprefix [0]{URL }%
\providecommand \Eprint[0]{\href }%
\@ifxundefined \urlstyle {%
  \providecommand \doi [1]{doi:\discretionary{}{}{}#1}%
}{%
  \providecommand \doi [0]{doi:\discretionary{}{}{}\begingroup
  \urlstyle{rm}\Url }%
}%
\providecommand \doibase [0]{http://dx.doi.org/}%
\providecommand \Doi[1]{\href{\doibase#1}}%
\providecommand \bibAnnote [3]{%
  \BibitemShut{#1}%
  \begin{quotation}\noindent
    \textsc{Key:}\ #2\\\textsc{Annotation:}\ #3%
  \end{quotation}%
}%
\providecommand \bibAnnoteFile [2]{%
  \IfFileExists{#2}{\bibAnnote {#1} {#2} {\input{#2}}}{}%
}%
\providecommand \typeout [0]{\immediate \write \m@ne }%
\providecommand \selectlanguage [0]{\@gobble}%
\providecommand \bibinfo [0]{\@secondoftwo}%
\providecommand \bibfield [0]{\@secondoftwo}%
\providecommand \translation [1]{[#1]}%
\providecommand \BibitemOpen[0]{}%
\providecommand \bibitemStop [0]{}%
\providecommand \bibitemNoStop [0]{.\EOS\space}%
\providecommand \EOS [0]{\spacefactor3000\relax}%
\providecommand \BibitemShut [1]{\csname bibitem#1\endcsname}%
%</preamble>
\bibitem{Tsui-PhysRevLett.48.1559}%
  \BibitemOpen
  \bibfield{author}{%
  \bibinfo {author} {\bibfnamefont{D.~C.}\ \bibnamefont{Tsui}}, \bibinfo
  {author} {\bibfnamefont{H.~L.}\ \bibnamefont{Stormer}},\ and\ \bibinfo
  {author} {\bibfnamefont{A.~C.}\ \bibnamefont{Gossard}},\ }%
  \bibfield{journal}{%
  \Doi{10.1103/PhysRevLett.48.1559}{\bibinfo {journal} {Phys. Rev. Lett.}}\ }%
  \textbf{\bibinfo {volume} {48}},\ \bibinfo {pages} {1559} (\bibinfo {month}
  {May}\ \bibinfo {year} {1982})%
  \bibAnnoteFile{NoStop}{Tsui-PhysRevLett.48.1559}%
\bibitem{Fubini-1991MPLA....6..347F}%
  \BibitemOpen
  \bibfield{author}{%
  \bibinfo {author} {\bibfnamefont{S.}~\bibnamefont{{Fubini}}},\ }%
  \bibfield{journal}{%
  \Doi{10.1142/S0217732391000336}{\bibinfo {journal} {Modern Physics Letters
  A}}\ }%
  \textbf{\bibinfo {volume} {6}},\ \bibinfo {pages} {347} (\bibinfo {year}
  {1991})%
  \bibAnnoteFile{NoStop}{Fubini-1991MPLA....6..347F}%
\bibitem{Moore1991362}%
  \BibitemOpen
  \bibfield{author}{%
  \bibinfo {author} {\bibfnamefont{G.}~\bibnamefont{Moore}}\ and\ \bibinfo
  {author} {\bibfnamefont{N.}~\bibnamefont{Read}},\ }%
  \bibfield{journal}{%
  \Doi{10.1016/0550-3213(91)90407-O}{\bibinfo {journal} {Nuclear Physics B}}\
  }%
  \textbf{\bibinfo {volume} {360}},\ \bibinfo {pages} {362 } (\bibinfo {year}
  {1991}),\ ISSN \bibinfo {issn} {0550-3213}%
  \bibAnnoteFile{NoStop}{Moore1991362}%
\bibitem{Laughlin-PhysRevLett.50.1395}%
  \BibitemOpen
  \bibfield{author}{%
  \bibinfo {author} {\bibfnamefont{R.~B.}\ \bibnamefont{Laughlin}},\ }%
  \bibfield{journal}{%
  \Doi{10.1103/PhysRevLett.50.1395}{\bibinfo {journal} {Phys. Rev. Lett.}}\ }%
  \textbf{\bibinfo {volume} {50}},\ \bibinfo {pages} {1395} (\bibinfo {month}
  {May}\ \bibinfo {year} {1983})%
  \bibAnnoteFile{NoStop}{Laughlin-PhysRevLett.50.1395}%
\bibitem{Read-PhysRevB.59.8084}%
  \BibitemOpen
  \bibfield{author}{%
  \bibinfo {author} {\bibfnamefont{N.}~\bibnamefont{Read}}\ and\ \bibinfo
  {author} {\bibfnamefont{E.}~\bibnamefont{Rezayi}},\ }%
  \bibfield{journal}{%
  \Doi{10.1103/PhysRevB.59.8084}{\bibinfo {journal} {Phys. Rev. B}}\ }%
  \textbf{\bibinfo {volume} {59}},\ \bibinfo {pages} {8084} (\bibinfo {month}
  {Mar}\ \bibinfo {year} {1999})%
  \bibAnnoteFile{NoStop}{Read-PhysRevB.59.8084}%
\bibitem{Bonderson-PhysRevB.83.075303}%
  \BibitemOpen
  \bibfield{author}{%
  \bibinfo {author} {\bibfnamefont{P.}~\bibnamefont{Bonderson}}, \bibinfo
  {author} {\bibfnamefont{V.}~\bibnamefont{Gurarie}},\ and\ \bibinfo {author}
  {\bibfnamefont{C.}~\bibnamefont{Nayak}},\ }%
  \bibfield{journal}{%
  \Doi{10.1103/PhysRevB.83.075303}{\bibinfo {journal} {Phys. Rev. B}}\ }%
  \textbf{\bibinfo {volume} {83}},\ \bibinfo {pages} {075303} (\bibinfo {month}
  {Feb}\ \bibinfo {year} {2011})%
  \bibAnnoteFile{NoStop}{Bonderson-PhysRevB.83.075303}%
\bibitem{Haldane-PhysRevLett.51.605}%
  \BibitemOpen
  \bibfield{author}{%
  \bibinfo {author} {\bibfnamefont{F.~D.~M.}\ \bibnamefont{Haldane}},\ }%
  \bibfield{journal}{%
  \Doi{10.1103/PhysRevLett.51.605}{\bibinfo {journal} {Phys. Rev. Lett.}}\ }%
  \textbf{\bibinfo {volume} {51}},\ \bibinfo {pages} {605} (\bibinfo {month}
  {Aug}\ \bibinfo {year} {1983})%
  \bibAnnoteFile{NoStop}{Haldane-PhysRevLett.51.605}%
\bibitem{Jain:1989p294}%
  \BibitemOpen
  \bibfield{author}{%
  \bibinfo {author} {\bibfnamefont{J.~K.}\ \bibnamefont{Jain}},\ }%
  \bibfield{journal}{%
  \bibinfo {journal} {Phys. Rev. Lett.}\ }%
  \textbf{\bibinfo {volume} {63}},\ \bibinfo {pages} {199} (\bibinfo {year}
  {1989})%
  \bibAnnoteFile{NoStop}{Jain:1989p294}%
\bibitem{Bergholtz-PhysRevB.77.155308}%
  \BibitemOpen
  \bibfield{author}{%
  \bibinfo {author} {\bibfnamefont{E.~J.}\ \bibnamefont{Bergholtz}}\ and\
  \bibinfo {author} {\bibfnamefont{A.}~\bibnamefont{Karlhede}},\ }%
  \bibfield{journal}{%
  \Doi{10.1103/PhysRevB.77.155308}{\bibinfo {journal} {Phys. Rev. B}}\ }%
  \textbf{\bibinfo {volume} {77}},\ \bibinfo {pages} {155308} (\bibinfo {month}
  {Apr}\ \bibinfo {year} {2008})%
  \bibAnnoteFile{NoStop}{Bergholtz-PhysRevB.77.155308}%
\bibitem{Bergholtz-PhysRevLett.99.256803}%
  \BibitemOpen
  \bibfield{author}{%
  \bibinfo {author} {\bibfnamefont{E.~J.}\ \bibnamefont{Bergholtz}}, \bibinfo
  {author} {\bibfnamefont{T.~H.}\ \bibnamefont{Hansson}}, \bibinfo {author}
  {\bibfnamefont{M.}~\bibnamefont{Hermanns}},\ and\ \bibinfo {author}
  {\bibfnamefont{A.}~\bibnamefont{Karlhede}},\ }%
  \bibfield{journal}{%
  \Doi{10.1103/PhysRevLett.99.256803}{\bibinfo {journal} {Phys. Rev. Lett.}}\
  }%
  \textbf{\bibinfo {volume} {99}},\ \bibinfo {pages} {256803} (\bibinfo {month}
  {Dec}\ \bibinfo {year} {2007})%
  \bibAnnoteFile{NoStop}{Bergholtz-PhysRevLett.99.256803}%
\bibitem{Seidel-PhysRevLett.95.266405}%
  \BibitemOpen
  \bibfield{author}{%
  \bibinfo {author} {\bibfnamefont{A.}~\bibnamefont{Seidel}}, \bibinfo {author}
  {\bibfnamefont{H.}~\bibnamefont{Fu}}, \bibinfo {author}
  {\bibfnamefont{D.-H.}\ \bibnamefont{Lee}}, \bibinfo {author}
  {\bibfnamefont{J.~M.}\ \bibnamefont{Leinaas}},\ and\ \bibinfo {author}
  {\bibfnamefont{J.}~\bibnamefont{Moore}},\ }%
  \bibfield{journal}{%
  \Doi{10.1103/PhysRevLett.95.266405}{\bibinfo {journal} {Phys. Rev. Lett.}}\
  }%
  \textbf{\bibinfo {volume} {95}},\ \bibinfo {pages} {266405} (\bibinfo {month}
  {Dec}\ \bibinfo {year} {2005})%
  \bibAnnoteFile{NoStop}{Seidel-PhysRevLett.95.266405}%
\bibitem{chak}%
  \BibitemOpen
  \bibfield{author}{%
  \bibinfo {author} {\bibfnamefont{T.}~\bibnamefont{Chakraborty}}\ and\
  \bibinfo {author} {\bibfnamefont{P.}~\bibnamefont{Pietilainen}},\ }%
  \emph{\bibinfo {title} {The Quantum {H}all Effects}}\ (\bibinfo {publisher}
  {Springer-Verlag, New York, Berlin, Heidelberg, Tokyo},\ \bibinfo {year}
  {1995})%
  \bibAnnoteFile{NoStop}{chak}%
\bibitem{Tao-PhysRevB.28.1142}%
  \BibitemOpen
  \bibfield{author}{%
  \bibinfo {author} {\bibfnamefont{R.}~\bibnamefont{Tao}}\ and\ \bibinfo
  {author} {\bibfnamefont{D.~J.}\ \bibnamefont{Thouless}},\ }%
  \bibfield{journal}{%
  \Doi{10.1103/PhysRevB.28.1142}{\bibinfo {journal} {Phys. Rev. B}}\ }%
  \textbf{\bibinfo {volume} {28}},\ \bibinfo {pages} {1142} (\bibinfo {month}
  {Jul}\ \bibinfo {year} {1983})%
  \bibAnnoteFile{NoStop}{Tao-PhysRevB.28.1142}%
\bibitem{Anderson-PhysRevB.28.2264}%
  \BibitemOpen
  \bibfield{author}{%
  \bibinfo {author} {\bibfnamefont{P.~W.}\ \bibnamefont{Anderson}},\ }%
  \bibfield{journal}{%
  \Doi{10.1103/PhysRevB.28.2264}{\bibinfo {journal} {Phys. Rev. B}}\ }%
  \textbf{\bibinfo {volume} {28}},\ \bibinfo {pages} {2264} (\bibinfo {month}
  {Aug}\ \bibinfo {year} {1983})%
  \bibAnnoteFile{NoStop}{Anderson-PhysRevB.28.2264}%
\bibitem{Su-PhysRevB.30.1069}%
  \BibitemOpen
  \bibfield{author}{%
  \bibinfo {author} {\bibfnamefont{W.~P.}\ \bibnamefont{Su}},\ }%
  \bibfield{journal}{%
  \Doi{10.1103/PhysRevB.30.1069}{\bibinfo {journal} {Phys. Rev. B}}\ }%
  \textbf{\bibinfo {volume} {30}},\ \bibinfo {pages} {1069} (\bibinfo {month}
  {Jul}\ \bibinfo {year} {1984})%
  \bibAnnoteFile{NoStop}{Su-PhysRevB.30.1069}%
\bibitem{Su-PhysRevB.32.2617}%
  \BibitemOpen
  \bibfield{author}{%
  \bibinfo {author} {\bibfnamefont{W.~P.}\ \bibnamefont{Su}},\ }%
  \bibfield{journal}{%
  \Doi{10.1103/PhysRevB.32.2617}{\bibinfo {journal} {Phys. Rev. B}}\ }%
  \textbf{\bibinfo {volume} {32}},\ \bibinfo {pages} {2617} (\bibinfo {month}
  {Aug}\ \bibinfo {year} {1985})%
  \bibAnnoteFile{NoStop}{Su-PhysRevB.32.2617}%
\bibitem{Chui-PhysRevB.32.8438}%
  \BibitemOpen
  \bibfield{author}{%
  \bibinfo {author} {\bibfnamefont{S.~T.}\ \bibnamefont{Chui}},\ }%
  \bibfield{journal}{%
  \Doi{10.1103/PhysRevB.32.8438}{\bibinfo {journal} {Phys. Rev. B}}\ }%
  \textbf{\bibinfo {volume} {32}},\ \bibinfo {pages} {8438} (\bibinfo {month}
  {Dec}\ \bibinfo {year} {1985})%
  \bibAnnoteFile{NoStop}{Chui-PhysRevB.32.8438}%
\bibitem{Chui-PhysRevLett.56.2395}%
  \BibitemOpen
  \bibfield{author}{%
  \bibinfo {author} {\bibfnamefont{S.~T.}\ \bibnamefont{Chui}},\ }%
  \bibfield{journal}{%
  \Doi{10.1103/PhysRevLett.56.2395}{\bibinfo {journal} {Phys. Rev. Lett.}}\ }%
  \textbf{\bibinfo {volume} {56}},\ \bibinfo {pages} {2395} (\bibinfo {month}
  {Jun}\ \bibinfo {year} {1986})%
  \bibAnnoteFile{NoStop}{Chui-PhysRevLett.56.2395}%
\bibitem{Bergholtz-PhysRevB.74.081308}%
  \BibitemOpen
  \bibfield{author}{%
  \bibinfo {author} {\bibfnamefont{E.~J.}\ \bibnamefont{Bergholtz}}, \bibinfo
  {author} {\bibfnamefont{J.}~\bibnamefont{Kailasvuori}}, \bibinfo {author}
  {\bibfnamefont{E.}~\bibnamefont{Wikberg}}, \bibinfo {author}
  {\bibfnamefont{T.~H.}\ \bibnamefont{Hansson}},\ and\ \bibinfo {author}
  {\bibfnamefont{A.}~\bibnamefont{Karlhede}},\ }%
  \bibfield{journal}{%
  \Doi{10.1103/PhysRevB.74.081308}{\bibinfo {journal} {Phys. Rev. B}}\ }%
  \textbf{\bibinfo {volume} {74}},\ \bibinfo {pages} {081308} (\bibinfo {month}
  {Aug}\ \bibinfo {year} {2006})%
  \bibAnnoteFile{NoStop}{Bergholtz-PhysRevB.74.081308}%
\bibitem{Seidel-PhysRevLett.97.056804}%
  \BibitemOpen
  \bibfield{author}{%
  \bibinfo {author} {\bibfnamefont{A.}~\bibnamefont{Seidel}}\ and\ \bibinfo
  {author} {\bibfnamefont{D.-H.}\ \bibnamefont{Lee}},\ }%
  \bibfield{journal}{%
  \Doi{10.1103/PhysRevLett.97.056804}{\bibinfo {journal} {Phys. Rev. Lett.}}\
  }%
  \textbf{\bibinfo {volume} {97}},\ \bibinfo {pages} {056804} (\bibinfo {month}
  {Aug}\ \bibinfo {year} {2006})%
  \bibAnnoteFile{NoStop}{Seidel-PhysRevLett.97.056804}%
\bibitem{Ardonne-1742-5468-2008-04-P04016}%
  \BibitemOpen
  \bibfield{author}{%
  \bibinfo {author} {\bibfnamefont{E.}~\bibnamefont{Ardonne}}, \bibinfo
  {author} {\bibfnamefont{E.~J.}\ \bibnamefont{Bergholtz}}, \bibinfo {author}
  {\bibfnamefont{J.}~\bibnamefont{Kailasvuori}},\ and\ \bibinfo {author}
  {\bibfnamefont{E.}~\bibnamefont{Wikberg}},\ }%
  \bibfield{journal}{%
  \bibinfo {journal} {Journal of Statistical Mechanics: Theory and Experiment}\
  }%
  \textbf{\bibinfo {volume} {2008}},\ \bibinfo {pages} {P04016} (\bibinfo
  {year} {2008})%
  \bibAnnoteFile{NoStop}{Ardonne-1742-5468-2008-04-P04016}%
\bibitem{Seidel-PhysRevLett.101.196802}%
  \BibitemOpen
  \bibfield{author}{%
  \bibinfo {author} {\bibfnamefont{A.}~\bibnamefont{Seidel}},\ }%
  \bibfield{journal}{%
  \Doi{10.1103/PhysRevLett.101.196802}{\bibinfo {journal} {Phys. Rev. Lett.}}\
  }%
  \textbf{\bibinfo {volume} {101}},\ \bibinfo {pages} {196802} (\bibinfo
  {month} {Nov}\ \bibinfo {year} {2008})%
  \bibAnnoteFile{NoStop}{Seidel-PhysRevLett.101.196802}%
\bibitem{Halperin83}%
  \BibitemOpen
  \bibfield{author}{%
  \bibinfo {author} {\bibfnamefont{B.~I.}\ \bibnamefont{Halperin}},\ }%
  \bibfield{journal}{%
  \bibinfo {journal} {Helv. Phys. Acta}\ }%
  \textbf{\bibinfo {volume} {56}},\ \bibinfo {pages} {75} (\bibinfo {year}
  {1983})%
  \bibAnnoteFile{NoStop}{Halperin83}%
\bibitem{Haldane-PhysRevLett.60.956}%
  \BibitemOpen
  \bibfield{author}{%
  \bibinfo {author} {\bibfnamefont{F.~D.~M.}\ \bibnamefont{Haldane}}\ and\
  \bibinfo {author} {\bibfnamefont{E.~H.}\ \bibnamefont{Rezayi}},\ }%
  \bibfield{journal}{%
  \Doi{10.1103/PhysRevLett.60.956}{\bibinfo {journal} {Phys. Rev. Lett.}}\ }%
  \textbf{\bibinfo {volume} {60}},\ \bibinfo {pages} {956} (\bibinfo {month}
  {Mar}\ \bibinfo {year} {1988})%
  \bibAnnoteFile{NoStop}{Haldane-PhysRevLett.60.956}%
\bibitem{Seidel-PhysRevLett.101.036804}%
  \BibitemOpen
  \bibfield{author}{%
  \bibinfo {author} {\bibfnamefont{A.}~\bibnamefont{Seidel}}\ and\ \bibinfo
  {author} {\bibfnamefont{K.}~\bibnamefont{Yang}},\ }%
  \bibfield{journal}{%
  \Doi{10.1103/PhysRevLett.101.036804}{\bibinfo {journal} {Phys. Rev. Lett.}}\
  }%
  \textbf{\bibinfo {volume} {101}},\ \bibinfo {pages} {036804} (\bibinfo
  {month} {Jul}\ \bibinfo {year} {2008})%
  \bibAnnoteFile{NoStop}{Seidel-PhysRevLett.101.036804}%
\bibitem{Seidel-PhysRevB.84.085122}%
  \BibitemOpen
  \bibfield{author}{%
  \bibinfo {author} {\bibfnamefont{A.}~\bibnamefont{Seidel}}\ and\ \bibinfo
  {author} {\bibfnamefont{K.}~\bibnamefont{Yang}},\ }%
  \bibfield{journal}{%
  \Doi{10.1103/PhysRevB.84.085122}{\bibinfo {journal} {Phys. Rev. B}}\ }%
  \textbf{\bibinfo {volume} {84}},\ \bibinfo {pages} {085122} (\bibinfo {month}
  {Aug}\ \bibinfo {year} {2011})%
  \bibAnnoteFile{NoStop}{Seidel-PhysRevB.84.085122}%
\bibitem{Jansen-2012JMP....53l3306J}%
  \BibitemOpen
  \bibfield{author}{%
  \bibinfo {author} {\bibfnamefont{S.}~\bibnamefont{{Jansen}}},\ }%
  \bibfield{journal}{%
  \Doi{10.1063/1.4768250}{\bibinfo {journal} {Journal of Mathematical
  Physics}}\ }%
  \textbf{\bibinfo {volume} {53}},\ \bibinfo {pages} {123306} (\bibinfo {month}
  {Dec.}\ \bibinfo {year} {2012}),\
  \Eprint{http://arxiv.org/abs/1109.4022}{arXiv:1109.4022 [math-ph]}%
  \bibAnnoteFile{NoStop}{Jansen-2012JMP....53l3306J}%
\bibitem{Soule-PhysRevB.85.155116}%
  \BibitemOpen
  \bibfield{author}{%
  \bibinfo {author} {\bibfnamefont{P.}~\bibnamefont{Soul\'e}}\ and\ \bibinfo
  {author} {\bibfnamefont{T.}~\bibnamefont{Jolicoeur}},\ }%
  \bibfield{journal}{%
  \Doi{10.1103/PhysRevB.85.155116}{\bibinfo {journal} {Phys. Rev. B}}\ }%
  \textbf{\bibinfo {volume} {85}},\ \bibinfo {pages} {155116} (\bibinfo {month}
  {Apr}\ \bibinfo {year} {2012})%
  \bibAnnoteFile{NoStop}{Soule-PhysRevB.85.155116}%
\bibitem{Nakamura-PhysRevLett.109.016401}%
  \BibitemOpen
  \bibfield{author}{%
  \bibinfo {author} {\bibfnamefont{M.}~\bibnamefont{Nakamura}}, \bibinfo
  {author} {\bibfnamefont{Z.-Y.}\ \bibnamefont{Wang}},\ and\ \bibinfo {author}
  {\bibfnamefont{E.~J.}\ \bibnamefont{Bergholtz}},\ }%
  \bibfield{journal}{%
  \Doi{10.1103/PhysRevLett.109.016401}{\bibinfo {journal} {Phys. Rev. Lett.}}\
  }%
  \textbf{\bibinfo {volume} {109}},\ \bibinfo {pages} {016401} (\bibinfo
  {month} {Jul}\ \bibinfo {year} {2012})%
  \bibAnnoteFile{NoStop}{Nakamura-PhysRevLett.109.016401}%
\bibitem{Wang-PhysRevB.87.245119}%
  \BibitemOpen
  \bibfield{author}{%
  \bibinfo {author} {\bibfnamefont{Z.-Y.}\ \bibnamefont{Wang}}\ and\ \bibinfo
  {author} {\bibfnamefont{M.}~\bibnamefont{Nakamura}},\ }%
  \bibfield{journal}{%
  \Doi{10.1103/PhysRevB.87.245119}{\bibinfo {journal} {Phys. Rev. B}}\ }%
  \textbf{\bibinfo {volume} {87}},\ \bibinfo {pages} {245119} (\bibinfo {month}
  {Jun}\ \bibinfo {year} {2013})%
  \bibAnnoteFile{NoStop}{Wang-PhysRevB.87.245119}%
\bibitem{Bergholtz2011755}%
  \BibitemOpen
  \bibfield{author}{%
  \bibinfo {author} {\bibfnamefont{E.~J.}\ \bibnamefont{Bergholtz}}, \bibinfo
  {author} {\bibfnamefont{M.}~\bibnamefont{Nakamura}},\ and\ \bibinfo {author}
  {\bibfnamefont{J.}~\bibnamefont{Suorsa}},\ }%
  \bibfield{journal}{%
  \Doi{http://dx.doi.org/10.1016/j.physe.2010.07.044}{\bibinfo {journal}
  {Physica E: Low-dimensional Systems and Nanostructures}}\ }%
  \textbf{\bibinfo {volume} {43}},\ \bibinfo {pages} {755 } (\bibinfo {year}
  {2011}),\ ISSN \bibinfo {issn} {1386-9477}%
  \bibAnnoteFile{NoStop}{Bergholtz2011755}%
\bibitem{Wang-PhysRevB.86.155104}%
  \BibitemOpen
  \bibfield{author}{%
  \bibinfo {author} {\bibfnamefont{Z.-Y.}\ \bibnamefont{Wang}}, \bibinfo
  {author} {\bibfnamefont{S.}~\bibnamefont{Takayoshi}},\ and\ \bibinfo {author}
  {\bibfnamefont{M.}~\bibnamefont{Nakamura}},\ }%
  \bibfield{journal}{%
  \Doi{10.1103/PhysRevB.86.155104}{\bibinfo {journal} {Phys. Rev. B}}\ }%
  \textbf{\bibinfo {volume} {86}},\ \bibinfo {pages} {155104} (\bibinfo {month}
  {Oct}\ \bibinfo {year} {2012})%
  \bibAnnoteFile{NoStop}{Wang-PhysRevB.86.155104}%
\bibitem{Ortiz-PhysRevB.88.165303}%
  \BibitemOpen
  \bibfield{author}{%
  \bibinfo {author} {\bibfnamefont{G.}~\bibnamefont{Ortiz}}, \bibinfo {author}
  {\bibfnamefont{Z.}~\bibnamefont{Nussinov}}, \bibinfo {author}
  {\bibfnamefont{J.}~\bibnamefont{Dukelsky}},\ and\ \bibinfo {author}
  {\bibfnamefont{A.}~\bibnamefont{Seidel}},\ }%
  \bibfield{journal}{%
  \Doi{10.1103/PhysRevB.88.165303}{\bibinfo {journal} {Phys. Rev. B}}\ }%
  \textbf{\bibinfo {volume} {88}},\ \bibinfo {pages} {165303} (\bibinfo {month}
  {Oct}\ \bibinfo {year} {2013})%
  \bibAnnoteFile{NoStop}{Ortiz-PhysRevB.88.165303}%
\bibitem{Bernevig-2012arXiv1204.5682B}%
  \BibitemOpen
  \bibfield{author}{%
  \bibinfo {author} {\bibfnamefont{B.~A.}\ \bibnamefont{{Bernevig}}}\ and\
  \bibinfo {author} {\bibfnamefont{N.}~\bibnamefont{{Regnault}}},\ }%
  \bibfield{journal}{%
  \bibinfo {journal} {ArXiv e-prints}}%
   (\bibinfo {month} {Apr.}\ \bibinfo {year} {2012}),\
  \Eprint{http://arxiv.org/abs/1204.5682}{arXiv:1204.5682 [cond-mat.str-el]}%
  \bibAnnoteFile{NoStop}{Bernevig-2012arXiv1204.5682B}%
\bibitem{Budich-PhysRevB.88.035139}%
  \BibitemOpen
  \bibfield{author}{%
  \bibinfo {author} {\bibfnamefont{J.~C.}\ \bibnamefont{Budich}}\ and\ \bibinfo
  {author} {\bibfnamefont{E.}~\bibnamefont{Ardonne}},\ }%
  \bibfield{journal}{%
  \Doi{10.1103/PhysRevB.88.035139}{\bibinfo {journal} {Phys. Rev. B}}\ }%
  \textbf{\bibinfo {volume} {88}},\ \bibinfo {pages} {035139} (\bibinfo {month}
  {Jul}\ \bibinfo {year} {2013})%
  \bibAnnoteFile{NoStop}{Budich-PhysRevB.88.035139}%
\bibitem{thomale}%
  \BibitemOpen
  \bibfield{author}{%
  \bibinfo {author} {\bibfnamefont{C.~H.~L.}\ \bibnamefont{Lee}}, \bibinfo
  {author} {\bibfnamefont{R.}~\bibnamefont{Thomale}},\ and\ \bibinfo {author}
  {\bibfnamefont{X.-L.}\ \bibnamefont{Qi}},\ }%
  \bibfield{journal}{%
  \bibinfo {journal} {Phys. Rev. B}\ }%
  \textbf{\bibinfo {volume} {88}},\ \bibinfo {pages} {035101} (\bibinfo {month}
  {Jul}\ \bibinfo {year} {2013})%
  \bibAnnoteFile{NoStop}{thomale}%
\bibitem{Simon-PhysRevB.75.075318}%
  \BibitemOpen
  \bibfield{author}{%
  \bibinfo {author} {\bibfnamefont{S.~H.}\ \bibnamefont{Simon}}, \bibinfo
  {author} {\bibfnamefont{E.~H.}\ \bibnamefont{Rezayi}},\ and\ \bibinfo
  {author} {\bibfnamefont{N.~R.}\ \bibnamefont{Cooper}},\ }%
  \bibfield{journal}{%
  \Doi{10.1103/PhysRevB.75.075318}{\bibinfo {journal} {Phys. Rev. B}}\ }%
  \textbf{\bibinfo {volume} {75}},\ \bibinfo {pages} {075318} (\bibinfo {month}
  {Feb}\ \bibinfo {year} {2007})%
  \bibAnnoteFile{NoStop}{Simon-PhysRevB.75.075318}%
\bibitem{Read-PhysRevB.61.10267}%
  \BibitemOpen
  \bibfield{author}{%
  \bibinfo {author} {\bibfnamefont{N.}~\bibnamefont{Read}}\ and\ \bibinfo
  {author} {\bibfnamefont{D.}~\bibnamefont{Green}},\ }%
  \bibfield{journal}{%
  \Doi{10.1103/PhysRevB.61.10267}{\bibinfo {journal} {Phys. Rev. B}}\ }%
  \textbf{\bibinfo {volume} {61}},\ \bibinfo {pages} {10267} (\bibinfo {month}
  {Apr}\ \bibinfo {year} {2000})%
  \bibAnnoteFile{NoStop}{Read-PhysRevB.61.10267}%
\bibitem{green-10thesis}%
  \BibitemOpen
  \bibfield{author}{%
  \bibinfo {author} {\bibfnamefont{D.}~\bibnamefont{Green}},\ }%
  \bibfield{journal}{%
  \bibinfo {journal} {Ph.D. thesis, Yale University, New Haven}}%
   (\bibinfo {year} {2001}),\ \bibinfo {note} {arXiv:cond-mat/0202455}%
  \bibAnnoteFile{NoStop}{green-10thesis}%
\bibitem{Read-PhysRevB.54.16864}%
  \BibitemOpen
  \bibfield{author}{%
  \bibinfo {author} {\bibfnamefont{N.}~\bibnamefont{Read}}\ and\ \bibinfo
  {author} {\bibfnamefont{E.}~\bibnamefont{Rezayi}},\ }%
  \bibfield{journal}{%
  \Doi{10.1103/PhysRevB.54.16864}{\bibinfo {journal} {Phys. Rev. B}}\ }%
  \textbf{\bibinfo {volume} {54}},\ \bibinfo {pages} {16864} (\bibinfo {month}
  {Dec}\ \bibinfo {year} {1996})%
  \bibAnnoteFile{NoStop}{Read-PhysRevB.54.16864}%
\bibitem{yellow}%
  \BibitemOpen
  \bibfield{author}{%
  \bibinfo {author} {\bibfnamefont{F.}~\bibnamefont{Di~Francesco}}, \bibinfo
  {author} {\bibfnamefont{P.}~\bibnamefont{Mathieu}},\ and\ \bibinfo {author}
  {\bibfnamefont{D.}~\bibnamefont{S\'en\'echal}},\ }%
  \emph{\bibinfo {title} {Conformal field theory}}\ (\bibinfo {publisher}
  {Springer},\ \bibinfo {year} {1997})%
  \bibAnnoteFile{NoStop}{yellow}%
\bibitem{Read-PhysRevB.79.245304}%
  \BibitemOpen
  \bibfield{author}{%
  \bibinfo {author} {\bibfnamefont{N.}~\bibnamefont{Read}},\ }%
  \bibfield{journal}{%
  \Doi{10.1103/PhysRevB.79.245304}{\bibinfo {journal} {Phys. Rev. B}}\ }%
  \textbf{\bibinfo {volume} {79}},\ \bibinfo {pages} {245304} (\bibinfo {month}
  {Jun}\ \bibinfo {year} {2009})%
  \bibAnnoteFile{NoStop}{Read-PhysRevB.79.245304}%
\bibitem{jainbook}%
  \BibitemOpen
  \bibfield{author}{%
  \bibinfo {author} {\bibfnamefont{J.~K.}\ \bibnamefont{Jain}},\ }%
  \emph{\bibinfo {title} {Composite Fermions}}\ (\bibinfo {publisher}
  {Cambridge University Press},\ \bibinfo {year} {2007})%
  \bibAnnoteFile{NoStop}{jainbook}%
\bibitem{Herland-PhysRevB.85.024520}%
  \BibitemOpen
  \bibfield{author}{%
  \bibinfo {author} {\bibfnamefont{E.~V.}\ \bibnamefont{Herland}}, \bibinfo
  {author} {\bibfnamefont{E.}~\bibnamefont{Babaev}}, \bibinfo {author}
  {\bibfnamefont{P.}~\bibnamefont{Bonderson}}, \bibinfo {author}
  {\bibfnamefont{V.}~\bibnamefont{Gurarie}}, \bibinfo {author}
  {\bibfnamefont{C.}~\bibnamefont{Nayak}},\ and\ \bibinfo {author}
  {\bibfnamefont{A.}~\bibnamefont{Sudb\o{}}},\ }%
  \bibfield{journal}{%
  \Doi{10.1103/PhysRevB.85.024520}{\bibinfo {journal} {Phys. Rev. B}}\ }%
  \textbf{\bibinfo {volume} {85}},\ \bibinfo {pages} {024520} (\bibinfo {month}
  {Jan}\ \bibinfo {year} {2012})%
  \bibAnnoteFile{NoStop}{Herland-PhysRevB.85.024520}%
\bibitem{Herland-PhysRevB.87.075117}%
  \BibitemOpen
  \bibfield{author}{%
  \bibinfo {author} {\bibfnamefont{E.~V.}\ \bibnamefont{Herland}}, \bibinfo
  {author} {\bibfnamefont{E.}~\bibnamefont{Babaev}}, \bibinfo {author}
  {\bibfnamefont{P.}~\bibnamefont{Bonderson}}, \bibinfo {author}
  {\bibfnamefont{V.}~\bibnamefont{Gurarie}}, \bibinfo {author}
  {\bibfnamefont{C.}~\bibnamefont{Nayak}}, \bibinfo {author}
  {\bibfnamefont{L.}~\bibnamefont{Radzihovsky}},\ and\ \bibinfo {author}
  {\bibfnamefont{A.}~\bibnamefont{Sudb\o{}}},\ }%
  \bibfield{journal}{%
  \Doi{10.1103/PhysRevB.87.075117}{\bibinfo {journal} {Phys. Rev. B}}\ }%
  \textbf{\bibinfo {volume} {87}},\ \bibinfo {pages} {075117} (\bibinfo {month}
  {Feb}\ \bibinfo {year} {2013})%
  \bibAnnoteFile{NoStop}{Herland-PhysRevB.87.075117}%
\bibitem{Zaletel-PhysRevB.86.245305}%
  \BibitemOpen
  \bibfield{author}{%
  \bibinfo {author} {\bibfnamefont{M.~P.}\ \bibnamefont{Zaletel}}\ and\
  \bibinfo {author} {\bibfnamefont{R.~S.~K.}\ \bibnamefont{Mong}},\ }%
  \bibfield{journal}{%
  \Doi{10.1103/PhysRevB.86.245305}{\bibinfo {journal} {Phys. Rev. B}}\ }%
  \textbf{\bibinfo {volume} {86}},\ \bibinfo {pages} {245305} (\bibinfo {month}
  {Dec}\ \bibinfo {year} {2012})%
  \bibAnnoteFile{NoStop}{Zaletel-PhysRevB.86.245305}%
\bibitem{Estienne-PhysRevB.87.161112}%
  \BibitemOpen
  \bibfield{author}{%
  \bibinfo {author} {\bibfnamefont{B.}~\bibnamefont{Estienne}}, \bibinfo
  {author} {\bibfnamefont{Z.}~\bibnamefont{Papi\ifmmode~\acute{c}\else
  \'{c}\fi{}}}, \bibinfo {author} {\bibfnamefont{N.}~\bibnamefont{Regnault}},\
  and\ \bibinfo {author} {\bibfnamefont{B.~A.}\ \bibnamefont{Bernevig}},\ }%
  \bibfield{journal}{%
  \Doi{10.1103/PhysRevB.87.161112}{\bibinfo {journal} {Phys. Rev. B}}\ }%
  \textbf{\bibinfo {volume} {87}},\ \bibinfo {pages} {161112} (\bibinfo {month}
  {Apr}\ \bibinfo {year} {2013})%
  \bibAnnoteFile{NoStop}{Estienne-PhysRevB.87.161112}%
\bibitem{Haldane-PhysRevLett.55.2095}%
  \BibitemOpen
  \bibfield{author}{%
  \bibinfo {author} {\bibfnamefont{F.~D.~M.}\ \bibnamefont{Haldane}},\ }%
  \bibfield{journal}{%
  \Doi{10.1103/PhysRevLett.55.2095}{\bibinfo {journal} {Phys. Rev. Lett.}}\ }%
  \textbf{\bibinfo {volume} {55}},\ \bibinfo {pages} {2095} (\bibinfo {month}
  {Nov}\ \bibinfo {year} {1985})%
  \bibAnnoteFile{NoStop}{Haldane-PhysRevLett.55.2095}%
\bibitem{Chandran-PhysRevB.84.205136}%
  \BibitemOpen
  \bibfield{author}{%
  \bibinfo {author} {\bibfnamefont{A.}~\bibnamefont{Chandran}}, \bibinfo
  {author} {\bibfnamefont{M.}~\bibnamefont{Hermanns}}, \bibinfo {author}
  {\bibfnamefont{N.}~\bibnamefont{Regnault}},\ and\ \bibinfo {author}
  {\bibfnamefont{B.~A.}\ \bibnamefont{Bernevig}},\ }%
  \bibfield{journal}{%
  \Doi{10.1103/PhysRevB.84.205136}{\bibinfo {journal} {Phys. Rev. B}}\ }%
  \textbf{\bibinfo {volume} {84}},\ \bibinfo {pages} {205136} (\bibinfo {month}
  {Nov}\ \bibinfo {year} {2011})%
  \bibAnnoteFile{NoStop}{Chandran-PhysRevB.84.205136}%
\bibitem{Simon-PhysRevB.75.075317}%
  \BibitemOpen
  \bibfield{author}{%
  \bibinfo {author} {\bibfnamefont{S.~H.}\ \bibnamefont{Simon}}, \bibinfo
  {author} {\bibfnamefont{E.~H.}\ \bibnamefont{Rezayi}}, \bibinfo {author}
  {\bibfnamefont{N.~R.}\ \bibnamefont{Cooper}},\ and\ \bibinfo {author}
  {\bibfnamefont{I.}~\bibnamefont{Berdnikov}},\ }%
  \bibfield{journal}{%
  \Doi{10.1103/PhysRevB.75.075317}{\bibinfo {journal} {Phys. Rev. B}}\ }%
  \textbf{\bibinfo {volume} {75}},\ \bibinfo {pages} {075317} (\bibinfo {month}
  {Feb}\ \bibinfo {year} {2007})%
  \bibAnnoteFile{NoStop}{Simon-PhysRevB.75.075317}%
\bibitem{Regnault-PhysRevLett.103.016801}%
  \BibitemOpen
  \bibfield{author}{%
  \bibinfo {author} {\bibfnamefont{N.}~\bibnamefont{Regnault}}, \bibinfo
  {author} {\bibfnamefont{B.~A.}\ \bibnamefont{Bernevig}},\ and\ \bibinfo
  {author} {\bibfnamefont{F.~D.~M.}\ \bibnamefont{Haldane}},\ }%
  \bibfield{journal}{%
  \Doi{10.1103/PhysRevLett.103.016801}{\bibinfo {journal} {Phys. Rev. Lett.}}\
  }%
  \textbf{\bibinfo {volume} {103}},\ \bibinfo {pages} {016801} (\bibinfo
  {month} {Jun}\ \bibinfo {year} {2009})%
  \bibAnnoteFile{NoStop}{Regnault-PhysRevLett.103.016801}%
\bibitem{Milovanovic-PhysRevB.82.035316}%
  \BibitemOpen
  \bibfield{author}{%
  \bibinfo {author} {\bibfnamefont{M.~V.}\
  \bibnamefont{Milovanovi\ifmmode~\acute{c}\else \'{c}\fi{}}}\ and\ \bibinfo
  {author} {\bibfnamefont{Z.}~\bibnamefont{Papi\ifmmode~\acute{c}\else
  \'{c}\fi{}}},\ }%
  \bibfield{journal}{%
  \Doi{10.1103/PhysRevB.82.035316}{\bibinfo {journal} {Phys. Rev. B}}\ }%
  \textbf{\bibinfo {volume} {82}},\ \bibinfo {pages} {035316} (\bibinfo {month}
  {Jul}\ \bibinfo {year} {2010})%
  \bibAnnoteFile{NoStop}{Milovanovic-PhysRevB.82.035316}%
\bibitem{Milovanovic-PhysRevB.80.155324}%
  \BibitemOpen
  \bibfield{author}{%
  \bibinfo {author} {\bibfnamefont{M.~V.}\
  \bibnamefont{Milovanovi\ifmmode~\acute{c}\else \'{c}\fi{}}}, \bibinfo
  {author} {\bibfnamefont{T.}~\bibnamefont{Jolicoeur}},\ and\ \bibinfo {author}
  {\bibfnamefont{I.}~\bibnamefont{Vidanovi\ifmmode~\acute{c}\else
  \'{c}\fi{}}},\ }%
  \bibfield{journal}{%
  \Doi{10.1103/PhysRevB.80.155324}{\bibinfo {journal} {Phys. Rev. B}}\ }%
  \textbf{\bibinfo {volume} {80}},\ \bibinfo {pages} {155324} (\bibinfo {month}
  {Oct}\ \bibinfo {year} {2009})%
  \bibAnnoteFile{NoStop}{Milovanovic-PhysRevB.80.155324}%
\bibitem{Toke-PhysRevB.80.205301}%
  \BibitemOpen
  \bibfield{author}{%
  \bibinfo {author} {\bibfnamefont{C.}~\bibnamefont{T\ifmmode~\mbox{\H{o}}\else
  \H{o}\fi{}ke}}\ and\ \bibinfo {author} {\bibfnamefont{J.~K.}\
  \bibnamefont{Jain}},\ }%
  \bibfield{journal}{%
  \Doi{10.1103/PhysRevB.80.205301}{\bibinfo {journal} {Phys. Rev. B}}\ }%
  \textbf{\bibinfo {volume} {80}},\ \bibinfo {pages} {205301} (\bibinfo {month}
  {Nov}\ \bibinfo {year} {2009})%
  \bibAnnoteFile{NoStop}{Toke-PhysRevB.80.205301}%
\bibitem{Freedman-PhysRevB.85.045414}%
  \BibitemOpen
  \bibfield{author}{%
  \bibinfo {author} {\bibfnamefont{M.~H.}\ \bibnamefont{Freedman}}, \bibinfo
  {author} {\bibfnamefont{J.}~\bibnamefont{Gukelberger}}, \bibinfo {author}
  {\bibfnamefont{M.~B.}\ \bibnamefont{Hastings}}, \bibinfo {author}
  {\bibfnamefont{S.}~\bibnamefont{Trebst}}, \bibinfo {author}
  {\bibfnamefont{M.}~\bibnamefont{Troyer}},\ and\ \bibinfo {author}
  {\bibfnamefont{Z.}~\bibnamefont{Wang}},\ }%
  \bibfield{journal}{%
  \Doi{10.1103/PhysRevB.85.045414}{\bibinfo {journal} {Phys. Rev. B}}\ }%
  \textbf{\bibinfo {volume} {85}},\ \bibinfo {pages} {045414} (\bibinfo {month}
  {Jan}\ \bibinfo {year} {2012})%
  \bibAnnoteFile{NoStop}{Freedman-PhysRevB.85.045414}%
\bibitem{seidelunpub}%
  \BibitemOpen
  \bibfield{author}{%
  \bibinfo {author} {\bibfnamefont{A.}~\bibnamefont{Weerasinghe}}\ and\
  \bibinfo {author} {\bibfnamefont{A.}~\bibnamefont{Seidel}},\ }%
  \bibfield{journal}{%
  \bibinfo {journal} {arXiv:1406.6444}}%
   (\bibinfo {year} {2014})%
  \bibAnnoteFile{NoStop}{seidelunpub}%
\bibitem{Hermanns-PhysRevB.83.241302}%
  \BibitemOpen
  \bibfield{author}{%
  \bibinfo {author} {\bibfnamefont{M.}~\bibnamefont{Hermanns}}, \bibinfo
  {author} {\bibfnamefont{N.}~\bibnamefont{Regnault}}, \bibinfo {author}
  {\bibfnamefont{B.~A.}\ \bibnamefont{Bernevig}},\ and\ \bibinfo {author}
  {\bibfnamefont{E.}~\bibnamefont{Ardonne}},\ }%
  \bibfield{journal}{%
  \Doi{10.1103/PhysRevB.83.241302}{\bibinfo {journal} {Phys. Rev. B}}\ }%
  \textbf{\bibinfo {volume} {83}},\ \bibinfo {pages} {241302} (\bibinfo {month}
  {Jun}\ \bibinfo {year} {2011})%
  \bibAnnoteFile{NoStop}{Hermanns-PhysRevB.83.241302}%
\bibitem{jackson}%
  \BibitemOpen
  \bibfield{author}{%
  \bibinfo {author} {\bibfnamefont{T.}~\bibnamefont{Jackson}}, \bibinfo
  {author} {\bibfnamefont{N.}~\bibnamefont{Read}},\ and\ \bibinfo {author}
  {\bibfnamefont{S.~H.}\ \bibnamefont{Simon}},\ }%
  \bibfield{journal}{%
  \bibinfo {journal} {Phys. Rev. B}\ }%
  \textbf{\bibinfo {volume} {88}},\ \bibinfo {pages} {075313} (\bibinfo {year}
  {2013})%
  \bibAnnoteFile{NoStop}{jackson}%
\bibitem{gaffnian_new1}%
  \BibitemOpen
  \bibfield{author}{%
  \bibinfo {author} {\bibfnamefont{B.}~\bibnamefont{Estienne}}, \bibinfo
  {author} {\bibfnamefont{N.}~\bibnamefont{Regnault}},\ and\ \bibinfo {author}
  {\bibfnamefont{B.~A.}\ \bibnamefont{Bernevig}},\ }%
  \bibfield{journal}{%
  \bibinfo {journal} {arXiv:1406.6262}}%
   (\bibinfo {year} {2014})%
  \bibAnnoteFile{NoStop}{gaffnian_new1}%
\bibitem{gaffnian_new2}%
  \BibitemOpen
  \bibfield{author}{%
  \bibinfo {author} {\bibfnamefont{T.}~\bibnamefont{Jolicoeur}}, \bibinfo
  {author} {\bibfnamefont{T.}~\bibnamefont{Mizusaki}},\ and\ \bibinfo {author}
  {\bibfnamefont{P.}~\bibnamefont{Lecheminant}},\ }%
  \bibfield{journal}{%
  \bibinfo {journal} {arXiv:1406.5891}}%
   (\bibinfo {year} {2014})%
  \bibAnnoteFile{NoStop}{gaffnian_new2}%
\bibitem{sw}%
  \BibitemOpen
  \bibfield{author}{%
  \bibinfo {author} {\bibfnamefont{J.~R.}\ \bibnamefont{Schrieffer}}\ and\
  \bibinfo {author} {\bibfnamefont{P.~A.}\ \bibnamefont{Wolff}},\ }%
  \bibfield{journal}{%
  \bibinfo {journal} {Phys. Rev.}\ }%
  \textbf{\bibinfo {volume} {149}},\ \bibinfo {pages} {491} (\bibinfo {year}
  {1966})%
  \bibAnnoteFile{NoStop}{sw}%
\bibitem{cut}%
  \BibitemOpen
  \bibfield{author}{%
  \bibinfo {author} {\bibfnamefont{F.}~\bibnamefont{Wegner}},\ }%
  \bibfield{journal}{%
  \bibinfo {journal} {Ann. Physik}\ }%
  \textbf{\bibinfo {volume} {3}},\ \bibinfo {pages} {77} (\bibinfo {year}
  {1994})%
  \bibAnnoteFile{NoStop}{cut}%
\end{thebibliography}%

\end{document}